\documentclass[11pt,a4paper]{article}                           

\usepackage{a4}          % Remove before submission

\usepackage{amsmath}
\usepackage{amssymb}
\usepackage[usenames]{color} 
\usepackage{graphicx}
\usepackage{array}
\usepackage{hhline}
\usepackage{microtype}
\usepackage[bf]{caption}
\usepackage{color}
\usepackage[usenames,dvipsnames]{xcolor}
\usepackage{tcolorbox}
\usepackage{multirow}
\usepackage{cite}
\usepackage{bbold}
 \usepackage{tabu}

\usepackage{ifpdf}
\ifpdf
  \usepackage[pdftex,
    pdftitle={},
    pdfauthor={},
    pdfsubject={},
    bookmarksopen, bookmarksnumbered, bookmarksopenlevel=2]{hyperref}
\fi
\def\hybrid{\topmargin -20pt    \oddsidemargin 0pt
        \headheight 0pt \headsep 0pt
        \textwidth 6.25in       % A4 paper
        \textheight 9 in       % A4 paper
        \marginparwidth .875in
        \parskip 5pt plus 1pt 
          \jot = 1.5ex
   }
\hybrid
\numberwithin{equation}{section}
\numberwithin{table}{section}

\newcommand{\beq}{\begin{equation}}  \newcommand{\eeq}{\end{equation}}
\newcommand{\bal}{\begin{aligned}}   \newcommand{\eal}{\end{aligned}}
\newcommand{\bea}{\begin{eqnarray}}  \newcommand{\eea}{\end{eqnarray}}

\newcommand{\bmat}{\left(\begin{array}}
\newcommand{\emat}{\end{array}\right)}

\newcommand{\nn}{\nonumber}

\newcommand{\cC}{\mathcal{C}}

\newcommand{\cN}{\mathcal{N}}

\newcommand{\bq}{{\bf q}}
\newcommand{\bp}{{\bf p}}
\newcommand{\ba}{{\bf a}}
\newcommand{\bx}{{\bf x}}
\newcommand{\bw}{{\bf w}}
\newcommand{\bP}{{\bf \Pi}}

\newcommand{\be}{\begin{equation}}
\newcommand{\ee}{\end{equation}}

\begin{document}

\baselineskip=14pt
\parskip 5pt plus 1pt

\vspace*{-1.5cm}
\begin{flushright}    % Publication numbers
  {\small
  }
\end{flushright}

\vspace{2cm}
\begin{center}        % Main title
  {\LARGE Stability of BPS States and Weak Coupling Limits}
\end{center}

\vspace{0.5cm}
\begin{center}        % Authors
{\large  Eran Palti}
\end{center}

\vspace{0.15cm}
\begin{center}        % Institutes
\emph{Department of Physics, Ben-Gurion University of the Negev, Beer-Sheva 84105, Israel}
             \\[0.15cm]
 
\end{center}

\vspace{2cm}

%%%%%%%%%%%%%%%%%%%%%%%%%%%%%%%%%%%%%%%%%%%%%%%
%%%%%%%%%%%%%%%%%%%%%%%%%%%%%%%%%%%%%%%%%%%%%%%
%%%%%%%%%%%%%%%%%%%%%%%%%%%%%%%%%%%%%%%%%%%%%%%
%%%%%%%%%%%%%%%%%%%%%%%%%%%%%%%%%%%%%%%%%%%%%%%
%%%%%%%%%%%%%%%%%%%%%%%%%%%%%%%%%%%%%%%%%%%%%%%
%%%%%%%%%%%%%%%%%%%%%%%%%%%%%%%%%%%%%%%%%%%%%%%
%%%%%%%%%%%%%%%%%%%%%%%%%%%%%%%%%%%%%%%%%%%%%%%
%%%%%%%%%%%%%%%%%%%%%%%%%%%%%%%%%%%%%%%%%%%%%%%

\begin{abstract}
\noindent   
We study the stability and spectrum of BPS states in ${\cal N}=2$ supergravity. We find evidence, and prove for a large class of cases, that BPS stability exhibits a certain filtration which is partially independent of the value of the gauge couplings. Specifically, for any perturbative value of any gauge coupling $g \ll 1$, a BPS state can only decay to some constituents if those constituents do not become infinitely heavier than it in the vanishing coupling limit $g \rightarrow 0$. This stability filtration can be mathematically formulated in terms of the monodromy weight filtration of the limiting mixed Hodge structure associated to the vanishing coupling limit. We study various implications of the result for the Swampland program which aims to understand such weak-coupling limits, specifically regarding the nature and presence of an infinite tower of light charged BPS states.
\end{abstract}

\thispagestyle{empty}
\clearpage

\tableofcontents

\setcounter{page}{1}

%%%%%%%%%%%%%%%%%%%%%%%%%%%%%%%%%%%%%%%%%%%%%%%
%%%%%%%%%%%%%%%%%%%%%%%%%%%%%%%%%%%%%%%%%%%%%%%
%%%%%%%%%%%                 %%%%%%%%%%%%%%%%%%%
%%%%%%%%%%%  DOCUMENT BODY  %%%%%%%%%%%%%%%%%%%
%%%%%%%%%%%                 %%%%%%%%%%%%%%%%%%%
%%%%%%%%%%%%%%%%%%%%%%%%%%%%%%%%%%%%%%%%%%%%%%%
%%%%%%%%%%%%%%%%%%%%%%%%%%%%%%%%%%%%%%%%%%%%%%%
%%%%%%%%%%%%%%%%%%%%%%%%%%%%%%%%%%%%%%%%%%%%%%%

%\newpage

%%%%%%%%%%%%%%%%%%%%
\section{Introduction}
%%%%%%%%%%%%%%%%%%%%

In this paper we study the behaviour of the BPS spectrum of states in ${\cal N}=2$ supergravity as we vary a gauge coupling to zero $g \rightarrow 0$. It is known that there is a certain universal structure to any weak-coupling limit (which reflects the nilpotent orbit theorem \cite{schmid}), as was studied in detail in \cite{Grimm:2018ohb,Grimm:2018cpv}. This universal structure allows for a general study of such limits. The spectrum of BPS states in ${\cal N}=2$ theories has been the topic of much work (see for example \cite{Moore10pitplectures} for a review). It is interesting because the spectrum is well controlled, specifically it can only change upon crossing walls of marginal stability in the moduli space (or coupling space). How the spectrum changes upon crossing these walls is described by the wall-crossing formula \cite{Kontsevich:2008fj}. 

We are motivated to study this for a number of reasons, which are all related in some way to ideas that are part of the Swampland program \cite{Vafa:2005ui,Palti:2019pca,vanBeest:2021lhn}. The central theme is the expectation that as $g \rightarrow 0$ one should find an infinite tower of charged BPS states becoming massless. We would like to understand various aspects of these states. We outline below the Swampland-related motivations. However, the results we find are quite general and go beyond the Swampland theme, and so we outline them first. 

%The wall-crossing formula gives a fine-structured description of the decay and binding of BPS states upon crossing walls of marginal stability. There is a much coarser requirement which is that simply for a state to decay to some constituents, it must be at least as heavy as the sum of their masses. The mass of the BPS states depends on the moduli, or gauge couplings, and so this condition can be calculated at a given locus in moduli space. Indeed, this is the condition which defines the wall of marginal stability itself. 

Our findings are then as follows: we find evidence that whether a state can decay to some constituents depends not only on their masses at the point in coupling space where the decay occurs, but on their masses all the way to the weak-coupling limit. So find evidence for a certain filtration in the BPS spectrum which is (partially) independent of the value of the couplings:

{\it {\bf BPS Stability Filtration:} In any weakly-coupled region $g \ll 1$, a BPS state can only decay to constituents whose mass is not infinitely higher than it in the $g \rightarrow 0$ limit.}

We are able to prove this property for a large class (one of the three types) of weakly-coupled regions.\footnote{By weakly-coupled we mean that instanton-type effects, so exponential in the coupling, are suppressed.} We also go some way towards showing that it holds completely generally.

We find this result quite striking. As an example, it implies that an electrically charged BPS black hole with a mass of our galaxy cannot emit a dyonic sub-Planckian BPS particle (upon crossing a wall of marginal stability) because in the $g \rightarrow 0$ limit the dyon will be infinitely heavier than the black hole (which would transition to a particle). 

The BPS filtration has a nice mathematical formulation: it is that the stability of BPS states in any weakly-coupled region is determined by their weight under the Monodromy Weight Filtration of the limiting mixed Hodge structure associated to the $g \rightarrow 0$ limit.

\subsection{Motivation from the Swampland program}

In this part of the introduction we outline the motivation for studying the spectrum and decay of BPS states approaching weak-coupling limits in the context of the Swampland. See also \cite{Blumenhagen:2018nts,Lee:2018urn,Gonzalo:2018guu,Gonzalo:2020kke,Corvilain:2018lgw,Joshi:2019nzi,Gonzalo:2019gjp,Marchesano:2019ifh,Font:2019cxq,Lee:2019xtm,Grimm:2019wtx,Erkinger:2019umg,Heidenreich:2019zkl,Lee:2019wij,Baume:2019sry,DallAgata:2020ino,EnriquezRojo:2020hzi,Andriot:2020lea,Cecotti:2020rjq,Gendler:2020dfp,Lee:2020gvu,Xu:2020nlh,Lanza:2020qmt,Heidenreich:2020ptx,Klaewer:2020lfg,Bastian:2020egp,Cota:2020zse} for studies of BPS and charged states in weak-coupling limits in the Swampland context.

\subsubsection*{Black hole stability and the black hole to particle transition}

The Swampland program aims to understand constrains on effective theories coupled to gravity from their quantum gravity ultraviolet completion. One of the approaches to this question is to try to utilize black hole physics. The prototypical example is the Weak Gravity Conjecture (WGC), which postulates the existence of a particle whose mass is less than its charge \cite{ArkaniHamed:2006dz}
\be
m \leq g q M_p \;,
\ee   
where here $m$ is the mass of the particle, $q$ its integer quantized charge, and $g$ the gauge coupling of the gauge field under which it is charged. It was proposed in \cite{ArkaniHamed:2006dz} that this constraint may be related to the discharge of extremal black holes, a process which requires such a particle. This is an attempt to connect physics of states whose mass is above the Planck scale, such as black holes, to states whose mass is below the Planck scale, such as the WGC particle. Crossing the Planck scale in this sense attempts to capture the input of quantum gravity.\footnote{There are some arguments we may try to make about how quantum gravity should behave to try and connect super-Planckian and sub-Planckian physics. For example, starting with a sub-extremal black hole and letting it shed mass through Hawking radiation, but no charge, until it is a Planck-scale remnant. But it is difficult to understand the physics of any such remnants, and whether they are problematic or not.} 

One can try to understand crossing the Planck scale in a more explicit way as follows. Consider an extremal state with charge $Q$ such that for a certain value of the gauge coupling $g=g_i$, we have
\be
g_i Q \gg 1 \;.
\ee
We may let this state be an extremal black hole with a super-Planckian mass, so saturating the extremality bound
\be
M \geq g_i Q M_p \;.
\ee
Now vary the gauge coupling to $g=g_f$, say by moving along a moduli space, into the regime of 
\be
g_f Q \ll 1 \;.
\ee 
Now the same state can no longer be considered as an extremal black hole, but may be regarded as a particle. In practice, tracking the state (while maintaining extremality) through this variation of the gauge coupling requires knowing the microscopics of quantum gravity, or having sufficient supersymmetry. We will return to this point, but let us continue to think about that this means. In particular, a contemporary perspective on the magnetic WGC \cite{Heidenreich:2015nta,Andriolo:2018lvp,Grimm:2018ohb,Heidenreich:2017sim,Lee:2018urn,Lee:2019wij,Palti:2019pca,vanBeest:2021lhn} is that there should be an infinite tower of charged states whose mass scale is set by the gauge coupling
\be
m_{\infty} \sim g M_p \;,
\ee
where $m_{\infty}$ is the mass scale of an infinite tower of states. Strictly speaking we should not consider this tower extended beyond the Planck scale, but it is infinite in the sense that as $g \rightarrow 0$ there are an infinite number of states below the Planck scale in the tower. Now we see that the state with charge $Q$ should be part of this tower at $g=g_f$, while it should be a charged black hole at $g=g_i$. This leads to tension, because on the one hand we were considering that charged black holes should be unstable and decay, on the other hand, we want a tower of stable charged states. 

%This is an example of the type of questions and challenges that the Swampland program faces in trying to use aspects of quantum gravity which we understand macroscopically, such as black holes (or de Sitter space), to derive some microscopic physics. In this case,

 We would like to understand what happens to a state as we vary the gauge coupling
\be
g Q \gg 1 \rightarrow \;g Q \ll 1 \;. 
\label{gmatogmi}
\ee
There is a way to understand this transition if there is sufficient supersymmetry. For example, with ${\cal N}=4$ supersymmetry we can track such a transition. Indeed, if we replace the gauge coupling by the string coupling $g_s$, then the two regimes correspond to the different descriptions of branes in string theory
\be
g_s Q \gg 1 \mathrm{\;(Black\;Holes)}\rightarrow \;g_s Q \ll 1 \mathrm{\;(Branes)}\;. 
\label{gsbhdb}
\ee
Following this transition is precisely what allowed the extraction of the microstates of black holes in string theory \cite{Strominger:1996sh}. However, the case of ${\cal N}=4$ supersymmetry is not so interesting for the questions we are after, because there is no decay processes for BPS states upon variations of the gauge couplings (paths in moduli space). Geometrically, the counting of the D-brane states is topological. 

The case of ${\cal N}=2$ supersymmetry is an interesting compromise between the complete control (but stability) of ${\cal N}=4$, and the instability and generality of no supersymmetry. With ${\cal N}=2$ supersymmetry BPS states are not necessarily stable over paths in moduli space, so there is something to understand about the transition (\ref{gmatogmi}). On the other hand, we have some control over their stability and spectrum. In particular, it can be that a state at one point in moduli space has no charged particles it can decay to, while after moving to a different point it no longer is BPS and can, and does, decay to some constituent particles. We will consider here specifically the case of type IIB string theory compactified on a Calabi-Yau manifold. Geometrically, in this setting the BPS states are no longer topologically counted, but correspond to D3 branes wrapping special Lagrangian sub-manifolds. 

It is worth emphasising that the sense of stability and decay of BPS states is technically quite different to the picture of a charged black hole emitting a charged particle, as in the Weak Gravity Conjecture. At a given point in moduli space, BPS states are just stable and do not decay at all. Decays can only be induced by varying the moduli, or equivalently, the couplings. Even then, a state cannot decay to a constituent which is mutually local to it. So an electric black hole will not decay by emitting an electric particle. Nonetheless, the idea is that studying the interaction between stability of black holes, the spectrum of particles, and the black hole to particle transition, in a controlled setting, will shed light on the same type of physics in less supersymmetric settings. 

\subsubsection*{Population of the light tower of BPS states}

Our analysis is based on a very general understanding of the behaviour of the masses of charged BPS states near any weak-coupling limit, as developed in \cite{Grimm:2018ohb} (see, \cite{Grimm:2018cpv,Corvilain:2018lgw,Grimm:2019bey,Grimm:2019ixq,Bastian:2020egp,Bastian:2021eom,Gendler:2020dfp} for follow-up work). One universal feature is that we can split the states into two types. Those which become massless at infinite distance in field space, or as $g \rightarrow 0$, which we term electric states. And those which become infinitely massive in that limit, which we term dyonic states. The light tower of charged states is therefore composed of purely electric states. We are then concerned with the question of the spectrum, and stability, of electric states along variations in moduli space approaching weak coupling limits.\footnote{Of course, the tower of states also relates to that of the distance conjecture \cite{Ooguri:2006in,Klaewer:2016kiy}, but the focus on a charged tower naturally places this in the context of the weak gravity conjecture.}

It was shown in \cite{Grimm:2018ohb} that approaching any weak-coupling limit $g \rightarrow 0$ there exists an infinite number of charges such that their associated BPS states would become massless in the limit. However, it was not proven that those BPS states are actually in the spectrum.\footnote{An argument was made for certain limits which showed that if one state in the so-called monodromy orbit is populated, they all are. This argument made an assumption that states in the orbit do not undergo decay upon the monodromy path. Actually, this is simple to prove: it follows from the fact that the decay can only occur with decay products that are not mutually local. But all magnetically charged states are heavier than the electric ones in the tower approaching infinite distance. So the argument indeed holds.} We would like to show that the states are indeed populated.

How can we study the spectrum of BPS states in this setting? In regimes where there is a geometric picture, such as type IIA mirrors in the large volume regime, it is possible to study the spectrum directly. But these are only very special loci in the type IIB moduli space, and more generally there are no geometric tools that can be applied directly on the type IIB side. Another method is to use duality with the Heterotic string \cite{Lee:2018urn,Lee:2018spm,Lee:2019tst,Lee:2019xtm,Lee:2019wij,Lee:2020gvu,Lee:2020blx}. In this work, we will utilise a tool which builds on the relation between black holes and D-branes (\ref{gsbhdb}). In our setting, the string coupling $g_s$ is part of the hypermultiplet moduli space, while the gauge couplings and BPS state masses and stability are controlled by the vector-multiplet moduli space, and the two moduli spaces decouple exactly.\footnote{At least they do so at the two-derivative level which corresponds to infinite Calabi-Yau volume. But since the volume is also a modulus, which also sits in the hypermultiplet sector, this limit does not affect the vector-multiplet sector.} We therefore expect some controlled correspondence between the microscopic D-brane spectrum and black hole physics. 

Such a correspondence has been developed extensively. starting with the works \cite{Moore:1998pn,Denef:2000nb,Denef:2002ru}. The idea is to associate a given charge with a black hole (even if the charge is such that the mass of the state is smaller than the Planck mass), and study how the moduli behave around this black hole. The moduli fields follow so-called attractor flows from their values at infinity, into the attractor loci on the black hole horizon, which is fixed uniquely by the charges (it is in this sense that a charge can be associated with a black hole, so more precisely is associated with an attractor locus). Then the claim is: if the attractor locus is in a controlled region of moduli space, the charge is populated by a BPS state. If the locus is in an uncontrolled region of moduli space, or it diverges to infinity, then the result is inconclusive regarding the presence of a BPS state with that charge. 

A crucial aspect of the attractors/BPS states correspondence is that we are considering the spectrum of a given charge at the value of the moduli at spatial infinity of the associated black hole solution. We must then allow for multi-centre black hole solutions, such that the multiple black holes have charges which sum to the charge of interest. If those component black holes have physical attractor loci, then the BPS state of the total charge is in the spectrum. Such multi-centre black holes can be understood as attractor flows which split on loci of marginal stability, and continue as separate flows to the different centres. 

\begin{figure}[h]
\centering
 \includegraphics[width=1.0\textwidth]{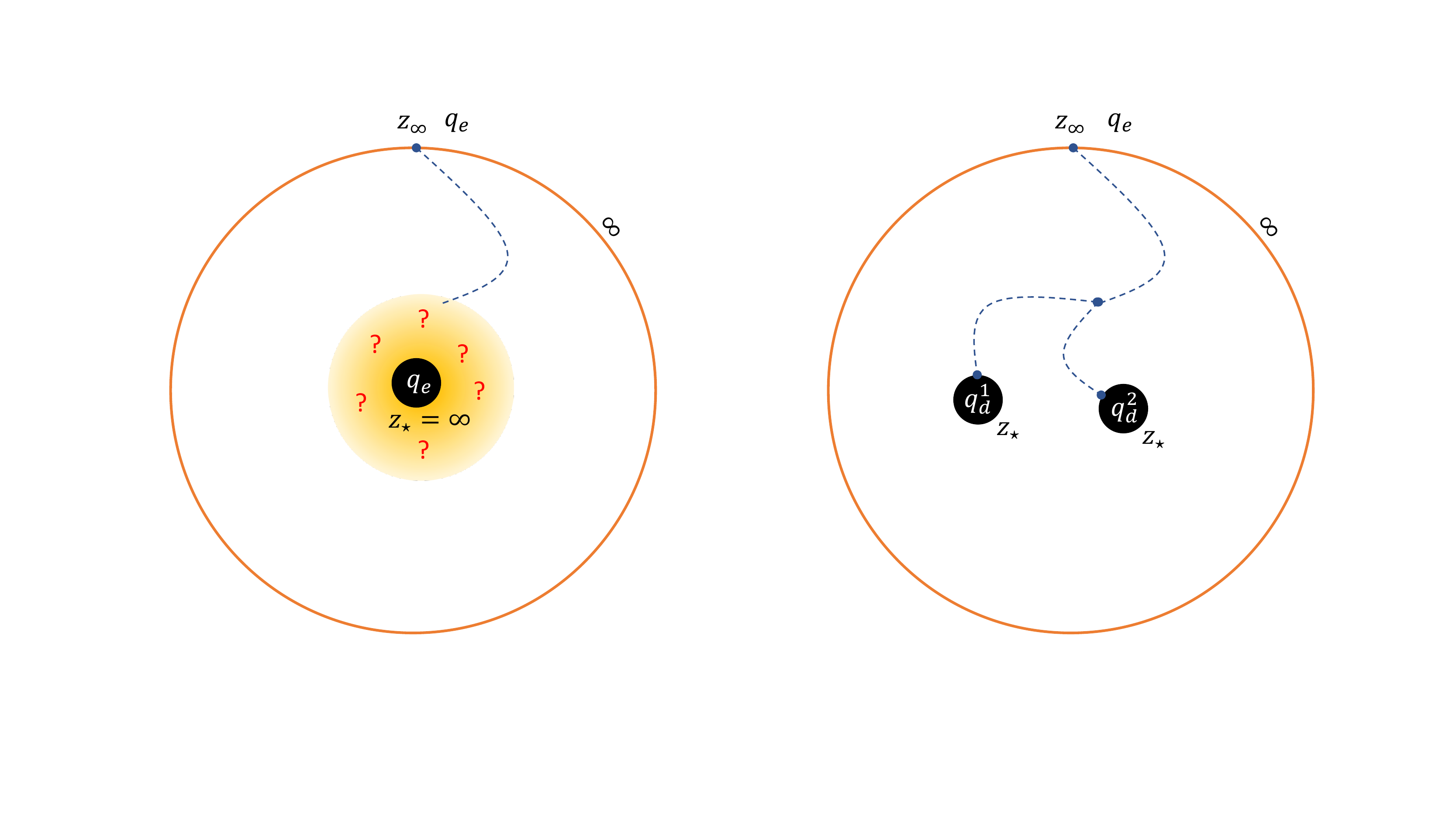}
\caption{Figure illustrating two attractor flows for a purely electric charge $\bq_e$. $z_{\infty}$ denotes the values of the moduli at spatial infinity, while $z_{\star}$ denotes their attractor values on the black hole horizon. The single-centred flow on the left is divergent and so not well-defined and cannot be used to deduce the presence of an associated BPS state. The split attractor flow on the right is well-defined and ends on two dyonic black holes with charges $\bq_d^1$ and $\bq_d^2$ at finite distance in moduli space.}
\label{fig:elesplit}
\end{figure}

If we try to probe the spectrum and stability of purely electrically charged states, those relevant for the tower, we encounter an immediate problem: the attractor flows of electric states diverge since it is possible to reduce their energy indefinitely by going to arbitrary weak gauge coupling. However, the existence of split attractor flows suggests a way around this issue: we can consider starting with a total electric charge at infinity, but then having a split attractor flow to a multi-centre black hole configuration where the constituent black holes are dyonic and have physical attractor loci. This is illustrated in figure \ref{fig:elesplit}. The existence of split flows corresponds to the decay of the electric BPS states into two dyonic states. We therefore are again motivated to understand BPS stability in the weakly-coupled regions of moduli space. 

\subsubsection*{The emergence proposal}

In \cite{Grimm:2018ohb} it was proposed that the weak coupling limit in ${\cal N}=2$ settings arises purely from integrating out the tower of light charged BPS states in that limit. This would be a controlled example of the more general emergence proposal \cite{Harlow:2015lma,Grimm:2018ohb,Palti:2019pca,Heidenreich:2017sim,Heidenreich:2018kpg}. If this is the case, then we should expect that the behaviour of BPS states, at least in weakly-coupled regions $g \ll 1$, would be controlled by the tower of states and so by the limit $g \rightarrow 0$. This is related also to the notion of moduli space holography \cite{Grimm:2020cda,Grimm:2021ikg}, where the whole moduli space is specified completely by its asymptotic limits (see also \cite{Cecotti:2020uek,Cecotti:2021cvv} for related ideas). So again we would like to understand how the BPS spectrum behaves in the bulk $g \ll 1$ with respect to its behaviour on the boundary $g \rightarrow 0$. 

In fact, the discussion above about how the spectrum of states can be understood from the existence of attractor flows already hints at some sort of emergence/holography. This is because it proposes that the spectrum of BPS states manifests non-locality in the moduli space. So the claim is that the presence of a BPS state at the values of the moduli which are taken at spatial infinity, is determined by the behaviour of the attractor loci on the black hole horizon, which are at a completely unrelated locus in moduli space. Such a notion of non-locality is naturally required for emergence/holography.

We will discuss the implications of the results for the motivation topics above in section \ref{sec:sum}. 

%%%%%%%%%%%%%%%%%%%%
\section{D3 branes as BPS states}
%%%%%%%%%%%%%%%%%%%%

In this work we are considering the setting of type IIB string theory on Calabi-Yau manifolds which gives an ${\cal N}=2$ supergravity. Within this we consider further only the vector-multiplet moduli space. The charged BPS states whose mass depends on the vector-multiplet (complex-structure) moduli are D3 branes wrapping 3-cycles in the Calabi-Yau. In this section, we discuss some properties of these states.   

%%%%%%%%%%%%%%%%%%%%
\subsection{${\cal N}=2$ BPS states}
\label{sec:d3n2bps}
%%%%%%%%%%%%%%%%%%%%

The moduli space of $\cN=2$ supersymmetric theories splits into two sectors, the vector multiplet moduli space and the hypermultiplet one. We will be solely concerned with the vector multiplet sector and denote by $n_V$ the number of vector multiplets. The moduli fields $z^i$, with $i=1,...,n_V$, control the mass of (half-)BPS states in the theory. The mass is set by the central charge $Z\left( \bq \right)$, which is a function of (integer quantised) charge vectors $\bq$. The charges are with respect to the gauge fields of the vector multiplets plus the graviphoton in the graviton multiplet. The moduli space and the central charge are controlled by the period vector $\bP\left(z^i \right)$, which has $2\left(n_V+1\right)$ entries that are holomorphic in the $z^i$. Specifically, 
\be
Z \left({\bq}\right) = e^{\frac{K}{2}} \bq^T \cdot \eta \cdot {\bf \Pi} \;.
\label{Zcc}
\ee
Here $\eta$ is a matrix defining a symplectic inner product, for example it can be taken as
\be
\eta = \left( \begin{array}{cc} 0 & \mathbb{1}_{n_V+1} \\  -\mathbb{1}_{n_V+1}  & 0  \end{array} \right) \;.
\label{etabas}
\ee
It is useful to denote the symplectic inner product as
\be
\left<\bq,\bp\right> \equiv \bq^T \cdot \eta \cdot \bp \;.
\ee
The Kahler potential, K, is itself determined by the period vector through 
\be
K = -\log i \left<\bP,\overline{\bP}\right> \;.
\ee

Compactifications of type IIB string theory on a Calabi-Yau manifold have BPS states corresponding to D3 branes wrapping three-cycles in the Calabi-Yau. The charge vector $\bq$ corresponds to the wrapping numbers of the $D3$ brane. While this is determined topologically, the question of whether there is an actual BPS state of a given charge is mapped to the existence of a special Lagrangian sub-manifold in the given homology class, which is not a topological invariant.

It is informative to look at the mass scale of the wrapped D3 branes in more detail. The expression for BPS states (\ref{Zcc}) is given in units of the four-dimensional (reduced) Planck mass $M_p$. This is related to the string scale $M_s$ as 
\be
M_s \sim \frac{g_s}{\sqrt{{\cal V}_s}} M_p \;. 
\ee 
Here $g_s$ is the string coupling, which is a field appearing in the hypermultiplet sector. ${\cal V}_s$ is the (dimensionless) volume of the Calabi-Yau in units of the string length, it is also a field which appears in the hypermultiplet sector. The mass of a $D3$ brane wrapped on a three-cycle $\cC$ takes the form
\be
M_{D3} \sim \frac{M_s}{g_s} {\cal V}_{\cC} \sim  \frac{M_s}{g_s} \sqrt{{\cal V}_s} Z\left(\bq\right) \sim Z\left(\bq\right) M_p \;,
\label{md3ms}
\ee
where here ${\cal V}_{\cC}$ denotes the volume in string units of $\cC$. 

We will be interested in this paper in limits in moduli space where $Z\left(\bq\right) \rightarrow 0$ or $Z\left(\bq\right) \rightarrow \infty$. We will refer to branes whose central charge goes to zero as light (electric) states, while branes whose central charge diverges will be referred to as heavy (dyonic) states. It is important to note that the expression (\ref{md3ms}) shows that these two ways to refer to the states should be interpreted with care. In particular, in the weakly-coupled supergravity regime we must have $ {\cal V}_{\cC} \gg 1$ and $g_s \ll 1$ which means that $M_{D3} \gg M_s$ irrespective of the value of $Z\left(\bq\right)$. Therefore, in that regime even the light $D3$ states are always heavier than the string states. On the other side, the heavy $D3$ states have $M_{D3} \gg M_p$, and so should not really be interpreted as particle states. Nonetheless, the presence of $\cN=2$ supersymmetry in the theory will allow us to probe certain aspects of the physics associated to both the light and heavy $D3$ states. It is these `protected' aspects of the $D3$ brane physics which will be the central elements in this paper. 

%%%%%%%%%%%%%%%%%%%%
\subsection{Decay and binding of BPS states}
\label{sec:decaybin}
%%%%%%%%%%%%%%%%%%%%

A given charge vector $\bq$ does not necessarily correspond to a stable BPS state in the theory. Further, the spectrum of stable BPS states depends on the point in moduli space at which it is evaluated. The changes to the spectrum of BPS states in the theory upon traversing paths in moduli space is controlled by Walls of Marginal Stability (WMS) where BPS states may either combine with other ones to form new stable bound states, or may become unstable and split into stable constituents. In this paper we will be primarily concerned with such process which involve only two constituents. 

Consider three BPS states $A$, $B$ and $C$ with respective charges $\bq_A$, $\bq_B$ and $\bq_{C}$. 
%The notation is that $\bar{A}$ is the anti-brane of brane $A$, and so $\bq_{\bar{A}}=-\bq_A$. 
We are interested in the process where the state $A$ decays into $B$ and $C$, denoted as $A \rightarrow B + C$. We therefore should impose charge conservation
\be
\bq_A = \bq_B + \bq_{C} \;.
\label{chargcon}
\ee
The central charge is a linear function of the charge, and therefore (\ref{chargcon}) implies
\be
Z\left(\bq_A\right) = Z\left(\bq_B\right) + Z\left(\bq_C\right) \;.
\label{cclinear}
\ee
The mass of the BPS states is
\be
M\left(\bq\right) = \left| Z\left(\bq\right)\right| \;.
\ee
We can then introduce the phase of the central charge $\alpha\left(\bq\right)$ as
\be
Z\left(\bq\right) = M\left(\bq\right) e^{i \alpha\left(\bq\right)} \;.
\label{gradedef}
\ee
%Note that for the purposes of the discussion in this section we may identify 
%\be
%\vp\left(\bq\right) \sim \vp\left(\bq\right) +2 \;.
%\ee
%In general, we will define the grades of a brane and anti-brane to be related as
%\be
%\vp\left(\bq_{\bar{A}}\right) = \vp\left(\bq_A\right) + 1 \;.  
%\label{graabb}
%\ee
From (\ref{cclinear}) we can write
\be
M\left(\bq_A\right)^2 = \left(M\left(\bq_B\right) + M\left(\bq_{C}\right) \right)^2 - 2 M\left(\bq_B\right) M\left(\bq_{C}\right) \big[1- \cos \left(\alpha\left(\bq_{B}\right)-\alpha\left(\bq_{C}\right) \right) \big]\;.
\label{trianineq}
\ee
Since the last term in (\ref{trianineq}) is negative we have $M\left(\bq_A\right) \leq M\left(\bq_B\right) + M\left(\bq_{C}\right)$ and therefore the decay $A \rightarrow B + C$ cannot occur unless the phases of $B$ and $C$ align. This alignment defines a Wall of Marginal Stability (WMS) for the decay $A \rightarrow B+C$,
\be
\mathrm{WMS}_{A \rightarrow B+C}\;:\; \alpha\left(\bq_{B}\right) = \alpha\left(\bq_{C}\right)  \;.
\label{wmsdec}
\ee
It is useful to note that (\ref{wmsdec}) implies also $\alpha\left(\bq_{A}\right)=\alpha\left(\bq_{B}\right)$.
However, the opposite direction is not quite true. Defining a wall of marginal stability by through the phases of $B$ and $A$ requires an additional condition on the masses
\be
\mathrm{WMS}_{A \rightarrow B+C}\;:\;\alpha\left(\bq_{A}\right) = \alpha\left(\bq_{B}\right) \;\mathrm{and}\; M\left(\bq_A\right) \geq M\left(\bq_{B}\right) \;.
\label{cond1}
\ee

The condition on the wall of marginal stability (\ref{wmsdec}) determines where such a decay process may occur, but it does not guarantee that it does. This depends on whether the constituent states are in the BPS spectrum themselves, and also on which direction we cross the wall. If crossing the wall in one direction we have the decay $A \rightarrow B+C$, then crossing he wall in the opposite direction must correspond to the reverse binding process $B + C \rightarrow A$. For small variations away from the WMS the condition on which process is occurring, decay or binding, is determined as \cite{Denef:2000nb,Aspinwall:2001dz,Denef:2002ru}
\bea
\mathrm{Decay}\;\left(A \rightarrow B+C\right)\;:\left<\bq_B,\bq_C\right>\left(\alpha\left(\bq_{B}\right) - \alpha\left(\bq_{C}\right)\right) < 0\;, \nn \\
\mathrm{Binding\;}\left(B + C \rightarrow A\right)\;:\; \left<\bq_B,\bq_C\right>\left(\alpha\left(\bq_{B}\right) - \alpha\left(\bq_{C}\right)\right) > 0 \;.
\eea
The microscopic physics associated to the decay or binding is the mass of open strings stretching between the branes. When the strings are tachyonic the branes form a bound state. Macroscopically, the condition corresponds to having a split attractor flow with a positive value for the splitting radius. The two pictures are related by varying the string coupling, as studied in \cite{Denef:2002ru}.

%%%%%%%%%%%%%%%%%%%%
\subsection{The BPS index and wall crossing formula}
\label{sec:BPSindex}
%%%%%%%%%%%%%%%%%%%%

A crucial aspect of the spectrum of BPS states is the degeneracy of states of a given charge. The deformation invariant quantity which provides a good measure of the degeneracy of states is the BPS index. The index is denoted as $\Omega\left(\bq,z\right)$ for a charge $\bq$ at a point in moduli space $z$. It receives a $+1$ from a massive hypermultiplet and $-2$ from a massive vector multiplet. The wrapped D3 branes give rise to hypermultiplets. 

In section \ref{sec:decaybin} we discussed walls of marginal stability. Upon crossing such walls, BPS states can bind or decay and correspondingly the BPS index can jump. Conversely, a change in the BPS index implies that the state has to cross some wall of marginal stability. The change in the BPS index upon crossing a wall is described in generality by the wall-crossing formula \cite{Kontsevich:2008fj}. We will not utilise the full details of the formula, and just give a feeling for it for sub-cases which were described in \cite{Denef:2007vg}. First we may consider the primitive decay process $\bq_A \rightarrow \bq_B + \bq_{C}$. By primitive we mean that $\bq_B$ and $\bq_{C}$ cannot be written as an integer multiple of some other quantized charge. In this case the change in the BPS index $\Omega\left(\bq,z\right)$ is given by \cite{Denef:2007vg}
\be
\Delta \Omega\left(\bq_A,z\right) = \left(-1\right)^{\left<\bq_B,\bq_{C}\right>-1}\left|\left<\bq_B,\bq_{C}\right> \right| \Omega\left(\bq_B,z_{ms}\right) \Omega\left(\bq_{C},z_{ms}\right) \;,
\ee
where $z_{ms}$ is the wall of marginal stability locus in moduli space. 

If we allow one of the charges to not be primitive, since the marginal stability locus corresponds to the alignment of the phases of $Z\left(\bq_B\right)$ and $Z\left(\bq_{C}\right)$, it is also a locus of marginal stability for the more general decay $\bq_A \rightarrow \bq_B + M_{AC}\;\bq_{C}$. The change in the BPS index for this decay is given by the generating function \cite{Denef:2007vg}
\be
\Omega\left(\bq_B\right)  +\sum_{X_{C\bar{A}} > 0}\Delta \Omega\left(\bq_B+M_{AC}\;\bq_{C}\right) l^{M_{AC}} = \Omega\left(\bq_B\right) \prod_{k>0} \left(1-(-1)^{k\left<\bq_B,\bq_{C}\right>}l^k \right)^{k\left|\left<\bq_B,\bq_{C}\right> \right|\Omega\left(k\;\bq_{C}\right) } \;.
\ee
Here $l$ is a dummy variable, and all the indices are evaluated on the locus of marginal stability. 

%%%%%%%%%%%%%%%%%%%%
\subsection{BPS states in Calabi-Yau manifolds}
\label{sec:BPSCY}
%%%%%%%%%%%%%%%%%%%%

It is useful to gain some intuition about the spectrum of BPS states in Calabi-Yau manifolds by looking at examples. We can consider the mirror IIA setting in the geometric supergravity regime, and look at D2-branes wrapping holomorphic curves. These can be calculated through mirror symmetry for example, or by counting directly holomorphic curves. The spectrum of such states for a two-parameter Calabi-Yau was calculated in \cite{Candelas:1993dm}. The CY is denoted $\mathbb{P}_4^{(1,1,2,2,2)}[8]$, which determines its construction as a complete intersection in weighted projective space. It has two homology classes for two-cycles, and associated Kahler moduli $v_t$ and $v_s$ which are part of vector multiplets with scalar components $t$ and $s$ denoted as
\be
t = b_t + i v_t \;,\;\; s = b_s + i v_s \;.
\ee  
In the large volume geometric regime, the Kahler potential for the Kahler moduli space is approximately given by
\be
e^{\frac{K}{2}} \sim \left( v_t^3 - 3 v_t v_s^2 - 2 v_s^3 \right)^{-\frac12} \;.
\ee
The mass of the D2 BPS states is given by the general central charge formula (\ref{Zcc}), except that now we are considering Kahler moduli of the CY. In this case it reads
\be
Z\left(j,k\right) = e^{\frac{K}{2}} \left[\frac{j}{2} t + \left(\frac{j}{2} - k\right) s \right] + {\cal O}\left(e^{2 \pi i t}, e^{2 \pi i s}\right) \;.
\ee
Here $j$ and $k$ are the two integers specifying the charges or wrapping numbers of the D2 branes on the two homology classes. The number of BPS states for $j$ and $k$ are shown in table 2.1. 
\begin{table}
\center
\begin{tabular}{|c||c|c|c|c|}
\hline
$j$ / $k$ & $0$ & $1$ & $2$ & $3$ \\
\hline
\hline	
0 & 0 & 4 & 0 & 0 \\
\hline	
1 & 640 & 640 & 0 & 0 \\
\hline	
2 & 10032 & 72224 & 10032 & 0 \\
\hline	
3 & 288384 & 7539200 & 7539200  & 288384 \\
\hline	
4 & 10979984  & 757561520  & 2346819520 & 757561520  \\
\hline	
5 & 495269504  & 74132328704 & 520834042880 & 520834042880 \\
\hline	
6 & 24945542832 & 7117563990784 & 7117563990784 & 212132862927264 \\
\hline	
\end{tabular}
\label{tab:bps2p}
\caption{Table taken from \cite{Candelas:1993dm} showing the number of BPS states for different wrapping numbers $j$ and $k$, in the Calabi-Yau manifold $\mathbb{P}_4^{(1,1,2,2,2)}[8]$.}	
\end{table}

There are of course much more advanced and complete calculations of BPS states in the geometric regime of type IIA, see for example \cite{Carta:2021sms} for the latest cutting edge. However, we are only interested in certain features of the BPS spectrum, which are universal to all examples. From table 2.1. we see observe:
\begin{enumerate}
\item Not all charges are populated by BPS states.
\item The populated charges have $k,j>0$ (or $k,j<0$ for the anti-branes).
\item If a charge $\bq$ supports a BPS state, this does not imply that $K \bq$ supports one, with $K$ some integer.
\item If a charge $K \bq$ supports a BPS state, with $K$ some integer, then so does the charge $\bq$.\footnote{In fact all charges $n \bq$ with $n \leq K$ support BPS states.}
\item There is a $\mathbb{Z}_2$ symmetry acting as $\left(j,k\right)\rightarrow \left(j,j-k \right)$
\end{enumerate}

The number of BPS states in a given homology class is counted by the holomorphic curves in the class. To gain some intuition for this it is simplest to consider holomorphic divisors. These are given by some holomorphic equations, say $f(z_i)=0$. The degree of the polynomial is associated to the homology class of the curve. The simplest such divisors to understand are those where the Calabi-Yau is embedded in a weighted projective space (as above) and the equations are in the ambient space coordinates which have certain weights. The weights of the polynomial determine the homology class. If we allow meromorphic functions, then we can span all the possible divisor classes. But a given class may not have any holomorphic representation, and in that case it will have a vanishing BPS count. This is the schematic understanding of property 1.

Property 2 seems straightforward to understand: if we have opposite signs for $k$ and $j$ then one wrapping direction will be anti-holomorphic with respect to the other direction, so the overall state cannot be BPS. In fact, this type of positivity property will play a crucial role in our analysis. In the wall-crossing formula it is related to the restriction to stable states under a certain quadratic form \cite{Kontsevich:2008fj}. 
%The existence of a basis with such positivity properties around a given point in moduli space is expected to be general, though this is not proven. 

Now consider property 3. From the picture of holomorphic divisors, having $\bq$ populated means we have some holomorphic $f(z_i)$ in that class. But then $f(z_i)^K$ is in the class $K \bq$ and so we may wonder how it cannot be populated. The point is that $f(z_i)^K$ is a factorised polynomial, and the BPS states are counted by non-factorizable holomorphic polynomials. So to have a representative we need some other monomial in the class $K \bq$, such that we can construct a non-factorized polynomial from the monomials. This does not always exist, and so we obtain property 3. 

Property 4 on the other hand says something like, if we have some holomorphic polynomial in class $K \bq$, then we must have one in $\bq$. This would be the case for example if the polynomial always contains a monomial $f(z_i)^K$. It is not clear why this should hold, but we can understand it by thinking about the origin of the homology classes which are empty of BPS states, which we denote as vanishing classes. Indeed, property 4 can also be stated as: If $\bq$ is a vanishing class, then so is $K \bq$. We observe in table 2.1 that there are infinite chains of vanishing classes. One way to understand these infinite chains is through the property 5. First note that holomorphicity is mapped to some positivity in the classes, $j,k \geq 0$, since we can only include positive powers of the coordinates. On the other hand, the $\mathbb{Z}_2$ symmetry maps any class with $k > j$ to a negative class, which cannot have holomorphic curves. Hence all classes with $k > j$ must be vanishing. So the infinite vanishing classes can be understood schematically as: there is one charge direction for which increasing the charge increases the negative powers of some coordinate. So that at some charge, we run out of holomorphic monomials and only have meromorphic ones. 

These infinite chains of vanishing classes are tied to the presence of a conifold locus in the mirror Calabi-Yau moduli space. The most explicit way to see this is that the $\mathbb{Z}_2$ symmetry is a sub-group of the monodromy group. In particular it corresponds to a combination of the conifold monodromy and another one \cite{Candelas:1993dm}. Traversing the monodromy path, and acting with the monodromy transformation on the charges is a gauge symmetry, see (\ref{monexch}). In general, this does not imply that the monodromy action on the charges is a symmetry of the spectrum. This is because somewhere along this path the BPS states can cross curves of marginal stability. If a state $\bq$ does not cross a wall of marginal stability upon a monodromy path, then the state $T \cdot \bq$ (with $T$ some matrix which is part of the monodromy group) must have the same number of BPS states \cite{Grimm:2018ohb,Denef:2007vg}. In the case when the monodromy action is a discrete finite one, say $\mathbb{Z}_N$, then the condition on not crossing a wall of marginal stability is automatic. This is because going around $N$ times with the same orientation must lead back to the original spectrum, but any wall of marginal stabilities would be crossed in the same direction upon traversing such a path, and so cannot switch between decay and binding walls as would be required. This is the reason why the $\mathbb{Z}_2$ monodromy subgroup is also a symmetry group of the spectrum. Its origin is the conifold monodromy, even though it is applied at large complex-structure (or more precisely at large volume in the IIA mirror).

It can be checked that property 4 above holds in many examples, possibly all known ones. For example, we found them to hold in all cases we looked at within the extensive database of \cite{Carta:2021sms}. We expect that it is general. 

%%%%%%%%%%%%%%%%%%%%
\subsection{Black hole attractors and BPS states}
\label{sec:entmon}
%%%%%%%%%%%%%%%%%%%%

Calculating the BPS index as a function of the moduli is too difficult. But there is a very useful way to probe the index by using the relation (\ref{gsbhdb}). Essentially, this is related to the fact that the entropy of a black hole is associated with the number of its microstates. In our case, the black holes are extremal supersymmetric black holes formed from D3 branes. The precise relation between the macroscopic entropy and the microscopic BPS index has been the subject of a large body of work (see \cite{DallAgata:2011zkh} for a review). We will primarily utilise the ideas first introduced in \cite{Denef:2000nb}. 

The entropy of black holes can be calculated through the horizon area, and counts the degeneracy of microstates with the black hole charge. Therefore, we can identify the macroscopic black hole entropy $S\left(\bq\right)$ with the (log of the) BPS index $\Omega\left(\bq\right)$. But the BPS index depends on the location in the moduli space, and therefore in this identification we must associate some point in moduli space to the black hole entropy. To understand what this means we need to introduce attractor flows. 

The charged black holes have two natural points in moduli space associated to them. The values of the moduli at the horizon of the black holes, these are the attractor values $z^i_{\star}$. And the values of the moduli at spatial infinity $z^i_{\infty}$. The central charge evaluated at these two values of the moduli corresponds to the black hole ADM mass and its entropy
\be
M_{ADM} = \left|Z\left(\bq,z^i_{\infty}\right) \right| \;,\;\;\; S = \pi \left|Z\left(\bq,z^i_{\star}\right)\right|^2 \;.
\label{scent}
\ee
The flow of the moduli between infinity and the horizon is described by the attractor mechanism \cite{Ferrara:1995ih,Ferrara:1997tw} (see \cite{DallAgata:2011zkh} for a review). For extremal black holes, the attractor moduli values are fixed by the charges of the black holes, and are independent of the moduli values at infinity. For a given charge of the black hole, our freedom amounts to the choice of the moduli values at infinity. The simplest cases, denoted double-extremal black holes in \cite{Kallosh:2006bt}, are when the moduli values at infinity are set equal to those on the horizon $z^i_{\infty}=z^i_{\star}$. In such situations there is no attractor flow at all. If instead we have $z^i_{\infty} \neq z^i_{\star}$ then we must have a non-trivial attractor flow. This flow is determined by the attractor flow equations, and is uniquely fixed. The attractor flow is monotonic in the central charge: throughout the flow the central charge is decreasing. The flow ends when the central charge is minimized
\be
\partial_i \left|Z\left(\bq\right)\right|_{\star} = 0 \;.
\label{attrlocu}
\ee

\begin{figure}
\centering
 \includegraphics[width=1.0\textwidth]{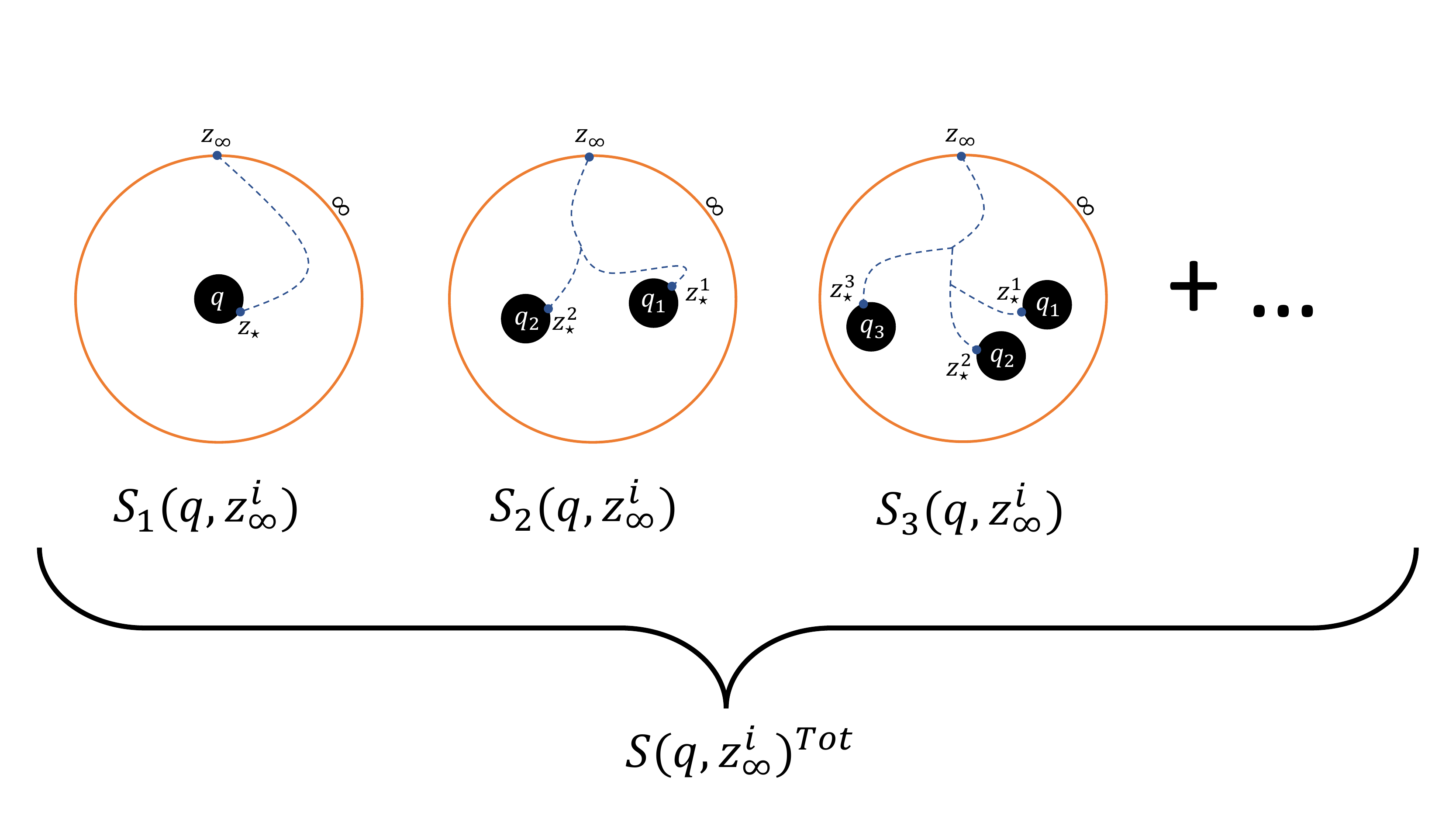}
\caption{Figure illustrating different black holes solutions which contribute to the total entropy $S\left(\bq,z^i_{\infty}\right)^{\mathrm{Tot}}$ associated with a fixed total charge and moduli values at infinity. Here infinity is denoted by the large circle, and the the moduli flows are shown as lines leading to attractor values on the black hole horizons.}
\label{fig:stot}
\end{figure}

Let us now return to the question of how to identify the black hole entropy with the BPS index. As explained in \cite{Denef:2000nb}, we should match the values of the moduli of the BPS index with the values of the moduli at spatial infinity, so we are calculating $\Omega\left(\bq,z^i_{\infty}\right)$. The index then receives contributions from two types of black holes: single-centered and multi-centered black holes. In the case of single-centered black holes we have a direct single attractor flow of the moduli from infinity to the black hole horizon. While in the case of multi-centered black holes the attractor flow splits on loci of marginal stability and each branch then leads to a different attractor point on different horizons of separate black holes. We can then associate a total entropy $S\left(\bq,z^i_{\infty}\right)^{\mathrm{Tot}}$  to a given charge and moduli values at infinity as the sum over the entropies of all the possible black hole solutions which have that (total) charge and asymptotic moduli values. It is this total entropy which can be identified with the (log of the) BPS index\footnote{There are some subtleties to do with this identification associated to potential black hole hair \cite{Banerjee:2009uk,Sen:2009vz}. We will ignore these subtleties as we do not expect that they can modify the analysis in this work.} 
\be
S\left(\bq,z^i_{\infty}\right)^{\mathrm{Tot}} \equiv \sum_{n}^{\mathrm{Black\;Holes}} S_n\left(\bq,z^i_{\infty}\right) = \log \Omega\left(\bq,z^i_{\infty}\right) \;.
\label{stotsum}
\ee
This is illustrated in figure \ref{fig:stot}.

%Note that if the moduli do not cross any loci of marginal stability along the attractor flow, then we only have the single-centered solution and so can identify $S\left(\bq,z^i_{\infty}\right)^{\mathrm{Tot}} = \pi \left|Z\left(\bq,z^i_{\star}\right)\right|^2$. In particular, this is the case for double-extremal black holes $z^i_{\infty}=z^i_{\star}$ since in such cases there is no attractor flow at all.

While it is rather natural to identify the entropy of single-centered black holes with the number of microscopic BPS states, the case of multi-centered black holes appears a little strange. After all, the microscopic BPS states are particles, while the multi-centered black holes have large spatial separation. The crucial point, as discussed for example in \cite{Denef:2002ru}, is the same one which relates the macroscopic entropy with microstates in a quantitative way: the string coupling $g_s$ does not correct the number of BPS states. The black hole description is valid at $g_s \left| \bq \right| \gg 1$, where $\left| \bq \right|$ is some appropriate measure of the amount of charge. While the microscopic BPS D-brane picture is valid for $g_s \left| \bq \right| \ll 1$. The limit $g_s \rightarrow 0$ is such that the string length becomes infinitely larger than the Planck length $\frac{l_s}{l_P} \rightarrow \infty$. The separation of the black hole centres is finite in Planck units, and therefore goes to zero in string units in the $g_s \rightarrow 0$ limit. In that limit, from the perspective of the microscopic string theory, they are a single object.

It is very important to emphasise that there are limitations to this macroscopic calculation of the BPS index. The attractor flows give only partial information about the microscopic BPS states. Certainly a well-controlled attractor flow and locus will contribute BPS states for a given charge, but an attractor flow which is not well-defined may or may not correspond to a state. In particular, this is the case for purely electric (or purely magnetic) charges where the attractor flow diverges due to the possibility of decreasing the energy density in the electric field arbitrarily by going to arbitrarily weak coupling. 

To illustrate this problem explicitly we can consider the spectrum of BPS states given in table 2.1. Since all the charges are purely electric, the attractor flows for all the black holes with those charges (or charges proportional to them) diverge to large volume. But the fact that some charges are populated with BPS states, while some are not, shows that we cannot deduce anything from these diverging flows. 

There is also an important connection to the analysis of the example spectrum in section \ref{sec:BPSCY} which we would like to utilise. The point is that if we accept the proposal that the spectrum of states is such that the population of a state of charge $K \bq$ implies the population of a state of charge $\bq$, then we can utilise split attractor flows as probes for towers of stable states. Specifically, we may construct a very massive electric state as a split flow into two dyonic states. If we find such flows (and we will not) then we deduce the population of a very massive super-Planckian electric state, but which would imply also that light sub-Planckian electric states are populated. Those, in particular, would be stable against any decay approaching weak coupling in moduli space, since they are lighter than any dyon. This way a split attractor flow can show the existence of a populated tower of states.

%%%%%%%%%%%%%%%%%%%%
\section{BPS states at infinite distance}
\label{sec:bpsinfdist}
%%%%%%%%%%%%%%%%%%%%

The structure of the vector multiplet moduli space exhibits certain universal features at infinite distance and weak coupling limits, which was developed in detail in \cite{Grimm:2018ohb,Grimm:2018cpv}. In this section we apply these results to determine the behaviour of the masses and phases of BPS states near those limits.

%%%%%%%%%%%%%%%%%%%%
\subsection{Moduli space near monodromies}
%%%%%%%%%%%%%%%%%%%%

The complex-structure, or vector multiplet, moduli space has certain singular loci where the period vector undergoes monodromies. If we denote the local coordinate about one of these singular loci as $z$, so that the locus is at $z=0$, then the monodromy action $\gamma$ corresponds to circling this locus
\be
\gamma \;:\; z \rightarrow z e^{r 2 \pi i}\;.
\label{mondpath}
\ee
The constant $r$ can take the values $\pm1$. The sign corresponds to going around the monodromy locus clockwise or anti-clockwise. This is a free choice at a global level, so we can choose to traverse all the monodromy loci together either clockwise or anti-clockwise, but once this choice is made there is a fixed global orientation which fixes the $\pm$ sign for each monodromy locus. This global orientation ensures that the product of all the monodromies gives the identity
\be
\prod_I \gamma_I = \mathbb{1} \;.
\ee
While this is in general important, it will not play a role in our analysis and so we henceforth set $r=+1$ for simplicity. 

Note that for moduli spaces with multiple moduli there are monodromy loci with complex codimension higher than one, and so have a number of monodromies associated to them. These settings will not be considered in detail in this paper, and we refer to, for example, \cite{Grimm:2018cpv} for an in-depth study of such monodromy loci. 

Under the monodromy the period vector transforms as
\be
\gamma \cdot \bP\left(z\right) = \bP \left(z e^{2 \pi i} \right) = T \cdot {\bf \Pi} \left(z \right) \;.
\label{monPdef}
\ee
Here $T$ is the monodromy matrix which has integer entries in an appropriate basis. Note that we will choose the coordinate $z$ in (\ref{monPdef}) such that it is the holomorphic coordinate of minimal power leading to a period vector which is holomorphic in it up to logarithmic factors (which then induce the monodromy). One can physically motivate this as saying (\ref{monPdef}) is the minimal rotation which leaves the physics invariant up to monodromy transformations.\footnote{In general, the monodromy matrix $T$ can be factorised as $T=T^{(o)} T^{(u)}$, where $T^{(o)}$ is a finite order matrix associated to a possible orbifolding of the moduli space, while $T^{(u)}$ is a unipotent matrix of infinite order. This will not be important for us because of this definition of $z$ as the minimal power.}

It is useful to introduce the matrix $N$ defined as
\be
T = e^{N} \;.
\ee
The unipotency of $T$ then translates to nilpotency of $N$ to a certain order. In particular, for Calabi-Yau threefolds $N^4=0$ for any monodromy locus. Indeed, an important property of a monodromy locus is the nilpotency order of its $N$. We define an integer $n$ such that 
\be
n \;: \; N^{n+1}=0 \;,\;\; N^n \neq 0 \;.
\label{ndef}
\ee 
The matrix $N$ controls the behaviour of the period vector. This is the Nilpotent Orbit Theorem \cite{schmid,Grimm:2018ohb} which states that 
\be
\bP\left(z\right) = \mathrm{Exp}\left[ N \frac{\log z}{2 \pi i}\right] \sum_{p=0}^{\infty} \ba_p z^p \;,
\label{not}
\ee
where the $\ba_p$ are constant (in $z$) vectors. We see that as $z \rightarrow 0$ the period vector is dominated by the $\log z$ parts with the $z^p$ parts contributing exponentially small corrections. This is true at least as long as $N \ba_0 \neq 0$. Indeed, an important quantity is the integer $d$ defined such that
\be
d \;: \; N^{d+1} \ba_0 =0 \;,\;\; N^d \ba_0 \neq 0 \;.
\label{ddef}
\ee
In particular, if $d=0$ then the monodromy locus is at finite distance in moduli space, like the conifold. While if $d >0$ then the locus is at infinite distance in moduli space. Conversely, any infinite distance locus in moduli space, which is also the case for any vanishing gauge coupling limit, has a monodromy with  $d>0$.

%%%%%%%%%%%%%%%%%%%%
\subsection{BPS states near monodromies}
\label{sec:bpsmodst}
%%%%%%%%%%%%%%%%%%%%

Combining the expression for the central charge (\ref{Zcc}) with that of the period vector (\ref{not}) we can obtain an approximate expression for the central charge. Consider the cases $d=1$ and $d=0$ where the relevant terms contributing to the central charge take the form 
\be
Z\left(\bq\right) = \frac{1}{\left| c \log \left|z\right| \right|^{\frac{1}{2}}} \left[ \left<\bq,\ba_0\right>  + \left<\bq,N\cdot\ba_0\right> \left( \frac{\log z}{2 \pi i} \right) + \;\left<\bq,\ba_1\right> z + ... \right] \;. 
\label{asympccgen}
\ee
Here $c$ is some unimportant constant. 
Let us consider the case $d=0$, which corresponds to the conifold locus. In this case, since $N \cdot\ba_0=0$, we have two types of states. Those whose charges satisfy  $\left<\bq,\ba_0\right> \neq 0$, which stay massive in the conifold limit $z \rightarrow 0$. And there are those where $\left<\bq,\ba_0\right> = 0$, which become massless in the conifold limit. 

Similarly, for the case $d=1$, there are states whose charges satisfy $\left<\bq,N\cdot\ba_0\right>\neq 0$ which become infinitely massive in the degeneration limit $z \rightarrow 0$. And states which satisfy $\left<\bq,N\cdot\ba_0\right> = 0$, that become massless in the degeneration limit. 

This type of decomposition into states that are massless and massive in the degeneration limit $z \rightarrow 0$ is general. We discuss aspects of this in section \ref{sec:decayfilt}, and refer to \cite{Grimm:2018ohb,Grimm:2018cpv} for more details. The result is that charges near the monodromy loci can be split into three types: dyonic charges labelled as $\bq_d$, magnetic charges labelled as $\bq_m$ and electric charges labelled as $\bq_e$. In the degeneration limit dyonic and magnetic states become infinitely massive, while electric states become massless. 

For $n=d=1$, they satisfy\footnote{It is worth noting that with these definitions of electric and magnetic states, they are not related by an electric-magnetic duality of the gauge coupling. In this sense, the magnetic states here should be thought of as dyonic also.}
\bea
\mathrm{(Massive)\;dyonic\;}&:&\;\left<\bq_d,N\cdot \ba_0\right> \neq 0 \;,\;\;\left<\bq_d,\ba_0\right> \neq 0 \;,  \nn \\
\mathrm{(Massive)\;magnetic\;}&:&\;\left<\bq_m,N\cdot \ba_0\right> \neq 0 \;,\;\;\left<\bq_m,\ba_0\right> \neq 0 \;, \nn \\
\mathrm{(Massless)\;electric\;}&:&\;\left<\bq_e,N\cdot \ba_0\right> = 0 \;,\;\;\left<\bq_e,\ba_0\right> \neq 0 \;.
\label{n1d1dyelma}
\eea
It is also the case that magnetic and electric states are mutually local
\be
\left<\bq^A_m,\bq^B_m\right> = \left<\bq^A_e,\bq^B_e\right> = 0 \;.
\label{eminner}
\ee

There is a natural action of the monodromy matrix $T$ on the charges $\bq$,
\be
\gamma \;:\;\; \bq \rightarrow T^{-1} \bq \;.
\label{monchmap}
\ee
%Note that in terms of the monodromy weight filtration, the monodromy matrix always shifts a charge by charges with lower weights. So if we denote the change to quantites induced by the monodromy as $\Delta_{\gamma}$ we have
%\be
%\Delta_{\gamma} \bq = \sum_{i=1}^{n} \left(-N\right)^i \cdot \bq \;.
%\ee
To see the action (\ref{monchmap}) we can track the monodromy action on the central charge 
\bea
Z\left( \bq,z \right) &=& e^{\frac{K}{2}} \left<\bq,\bP\left(z\right)\right> \nn \\
&\xrightarrow{\gamma}& e^{\frac{K}{2}} \left<\bq,\bP\left(ze^{2 \pi i}\right)\right> =  e^{\frac{K}{2}}\left<\bq,T \cdot \bP\left(z\right)\right> =  e^{\frac{K}{2}} \left<T^{-1} \cdot\bq, \bP\left(z\right)\right>\nn \\
&=&  Z\left( T^{-1} \cdot \bq,z \right)  \;.
\eea
The monodromy is a gauge transformation, which means are can always exchange the monodromy path on the moduli space for the action on the charges
\be
\left\{z \rightarrow z e^{2 \pi i} \right\} \leftrightarrow \left\{ \bq \rightarrow T^{-1} \cdot\bq \right\}\;.
\label{monexch}
\ee

Note that BPS states satisfy an identity (see, for example, \cite{Palti:2017elp})
\be
{\cal Q}\left(\bq\right)^2 = \left|Z\left(\bq\right)\right|^2 + g^{ij}D_iZ\left(\bq\right) \overline{D}_j \overline{Z}\left(\bq\right) \;.
\ee
Here ${\cal Q}\left(\bq\right)^2$ is a measure of the physical charge of the state, the precise form of which is not important but for a single $U(1)$ it is schematically like $g^2 \left|\bq\right|^2$ with $g$ the gauge coupling. From this we see, as expected, that the degeneration locus $\log \left|z\right| \rightarrow -\infty$ also corresponds to a vanishing gauge coupling (at least for the linear combination of gauge fields associated to the charge $\bq$). Indeed, generally, electric charges are with respect to electric gauge fields, all of which have a gauge coupling which behaves as \cite{Grimm:2018ohb}
\be
g \sim \left(\left|\log \left|z\right| \right|\right)^{-\frac{d}{2}} \;.
\label{gcou}
\ee

%%%%%%%%%%%%%%%%%%%%
\subsection{Walls of marginal stability at infinite distance}
\label{sec:wmsmon}
%%%%%%%%%%%%%%%%%%%%

In this section we study the structure of walls of marginal stability near infinite distance loci. We consider for simplicity the $d=1$ case, but the conclusions are general. 

Walls of Marginal Stability (WMS) are defined by the phase equalities (\ref{wmsdec}) and (\ref{cond1}). This is a real condition and therefore they trace out real dimension one lines in the moduli space. 
We will often work with a change of coordinates
\be
t = \frac{\log z}{2 \pi i} \;.
\label{ztot}
\ee
So that $\mathrm{Im\;}t \rightarrow +\infty$ is the degeneration limit, and the monodromy path $\gamma$ is $ \mathrm{Re\;}t \rightarrow \mathrm{Re\;}t + 1$. 

Let us consider electric states near infinite distances. From the central charge expression (\ref{asympccgen}), and the definitions (\ref{n1d1dyelma}), we see that electric states have central charges of the form
\be
Z\left(\bq\right) = \frac{1}{\left| c \;\mathrm{Im\;}t \right|^{\frac{1}{2}}} \left[ \left<\bq_e,\ba_0\right> + \left<\bq_e,\ba_1\right> e^{2 \pi i t} + ... \right] \;. 
\label{asymelec}
\ee
 We see that variations in moduli space do not change the phase of the central charge (up to exponentially small corrections). So approaching infinite distance, electric states have a static phase. Since the charge space of electric states is at least two (real) dimensional, the phases of electric states are static points distributed throughout the phase circle.
 
There are no walls of marginal stability (in the $t$ plane) for purely electric states to decay to other purely electric states when approaching infinite distance. So if they are present in the spectrum at some large $\mathrm{Im\;}t$, increasing $\mathrm{Im\;}t$ further (and varying $\mathrm{Re\;}t$ as we wish) will not lead to a decay to other purely electric states. Even without the phase, since the inner product of electric states with each other vanishes (\ref{eminner}), we see that they cannot decay to other electric states.
%\footnote{Note that electric and magnetic states are distinguished only up to exponentially large charges. More precisely, we see from (\ref{asymelec}) that the term $\left<\bq_e,\ba_1\right> e^{2 \pi i t} $ may compete with $\left<\bq_e,\ba_0\right>$ if we have some sufficiently large charges which induce an exponential difference between $\left<\bq_e,\ba_0\right>$ and $\left<\bq_e,\ba_1\right>$.}
% Charges which satisfy 
%  \be
%  \frac{\left<\bq_e,\ba_0\right>}{\left| \;\mathrm{Im\;}t \right|^{\frac{1}{2}}} \gg 1 \;,
%  \label{elebh}
%  \ee
%  and so which are associated with black holes not particles. 
% 

Magnetically charged states (which could be purely magnetic or dyonic) have a central charge such that all the terms in (\ref{asympccgen}) are generically non-vanishing. We can write this as
\be
Z\left(\bq\right) = \frac{\mathrm{Im\;}t}{\left| c \;\mathrm{Im\;}t  \right|^{\frac{1}{2}}} \left[i \left<\bq_m, N\cdot\ba_0\right> + \frac{\left<\bq,\ba_0\right>  + \left<\bq_m, N\cdot\ba_0\right> \mathrm{Re\;}t }{\mathrm{Im\;}t} + ... \right] \;,
\label{asympccgendyp}
\ee
where $\bq_m$ is the magnetic component of the dyonic charge. The variation of the phase under paths in moduli space is suppressed again but now only by $\mathrm{Im\;}t$ and not exponentially. This means that dyonic states can cross walls of marginal stability with respect to other dyonic states and also electric states. 

With respect to the stability of purely electric states, we see that purely electric states which satisfy 
\be
  \frac{\left<\bq_e,\ba_0\right>}{\left| \;\mathrm{Im\;}t \right|^{\frac{1}{2}}} \ll 1 \;,
  \label{elebhmc}
  \ee
are stable because they cannot decay to dyonic states as they are too light. So at infinite distance we have an infinite number of states that, if populated, would be stable \cite{Grimm:2018ohb}. But importantly, such states are sub-Planckian in mass, so they are not electric black holes. Black holes, so states which satisfy instead
  \be
  \frac{\left<\bq_e,\ba_0\right>}{\left| \;\mathrm{Im\;}t \right|^{\frac{1}{2}}} \gg 1 \;,
  \label{elebh}
  \ee
 can decay to dyonic states if those are present in the spectrum. 
  
It is simple to state the wall of marginal stability between two general dyonic states. Let us consider three charges $\bq^A = \bq^B + \bq^C$, and we are interested in a wall of marginal stability for the process 
\be
\bq^A \rightarrow \bq^B + \bq^C \;.
\ee
There are two ways to formulate the condition for the wall of marginal stability, corresponding to (\ref{wmsdec}) and (\ref{cond1}). 

In terms of the charges $\bq^B$ and $\bq^C$, as in (\ref{wmsdec}), we can write the condition for the wall of marginal stability as 
\be
\mathrm{Re\;} Z\left(\bq^B,t_{WMS}\right) \mathrm{Im\;} Z\left(\bq^C,t_{WMS}\right)  = \mathrm{Re\;} Z\left(\bq^C,t_{WMS}\right)  \mathrm{Im\;} Z\left(\bq^B,t_{WMS}\right)  \;,
\label{bcwmsZc1}
\ee
where additionally we must impose 
\be
\frac{\mathrm{Im\;} Z\left(\bq^B,t_{WMS}\right)}{\mathrm{Im\;} Z\left(\bq^C,t_{WMS}\right)}  > 0\;.
\label{bcwmsZc2}
\ee
The second condition ensures that (\ref{bcwmsZc1}) corresponds to alignment rather than anti-alignment. 

Here we have introduced the notation $t_{WMS}$, which refers to the value of $t$ on the wall of marginal stability. Note that this is a wall, so a one real parameter line in complex $t$ space, and so $t_{WMS}$ is a parameter. 

In terms of the charges $\bq^A$ and $\bq^B$, as in (\ref{cond1}), we can write the condition for the wall of marginal stability as
\be
\mathrm{Re\;} Z\left(\bq^A,t_{WMS}\right) \mathrm{Im\;} Z\left(\bq^B,t_{WMS}\right)  = \mathrm{Re\;} Z\left(\bq^B,t_{WMS}\right)  \mathrm{Im\;} Z\left(\bq^A,t_{WMS}\right)  \;,
\label{wmsZc1}
\ee
where additionally we must impose 
\be
\frac{Z\left(\bq^A,t_{WMS}\right)}{Z\left(\bq^B,t_{WMS}\right)} \geq 1\;.
\label{wmsZc2}
\ee
The condition (\ref{wmsZc2}) ensures that it is alignment, rather than anti-alignment, which is solving (\ref{wmsZc1}), as well as ensuring the second constraint in (\ref{cond1}).

%%%%%%%%%%%%%%%%%%%%
\subsection{Single-centre attractor flows at infinite distance}
\label{sec:attflnd1}
%%%%%%%%%%%%%%%%%%%%

In this section we study single-centre attractor flows near infinite distance loci corresponding to $n=d=1$ in the classification of (\ref{ddef}). We aim to determine which charges support attractor loci which are in a controlled region of moduli space. For the modulus $t$ associated to the infinite distance, this requirement on the attractor locus is
\be
\mathrm{Im\;}t_{\star} \gg 1 \;.
\label{attrimgt}
\ee
For the other moduli, which we will refer to as $z_i$, to clearly distinguish them from $t$, the condition can be stated generally that they should have finite attractor values within the bulk of the moduli space. This need not be near some infinite distance locus, or where any Nilpotent orbit expansion is manifest. Let us denote the space of such physical values of the moduli as ${\cal M}_{\mathrm{Phys}}$. We therefore require for a viable black hole to exist that (\ref{attrlocu}) has solutions such that
\be
z^i_{\star} \in {\cal M}_{\mathrm{Phys}} \;.
\label{zisphys}
\ee

The central charge is a restriction of (\ref{asympccgen}) and takes the form
\be
Z\left(\bq,t,z^i\right) = \frac{1}{\left| c\left(z^i\right) \mathrm{Im\;}t \right|^{\frac{1}{2}}} \left[ \left<\bq,\ba_0\left(z^i\right)\right>  + \left<\bq,N\cdot\ba_0\left(z^i\right)\right> t + ... \right] \;. 
\label{Ztd1n1}
\ee
Here we dropped the terms exponentially suppressed in $\mathrm{Im\;}t$, and have manifested the dependence on the other moduli $z_i$.
 The attractor locus for the modulus $t$ is then
\be
\bar{t}_{\star} = -\frac{\left<\bq,\left(\ba_0 \right)_{\star}\right>}{\left<\bq,N \cdot \left(\ba_0 \right)_{\star}\right>} \;,
\label{tstb}
\ee
where $\left(\ba_0 \right)_{\star}$ means evaluated on the attractor solution for the other fields $z^i = z^i_{\star}$. 
This gives
\be
\mathrm{Im\;}t_{\star} = \frac{\mathrm{Re\;}\left<\bq,N\cdot\left(\ba_0 \right)_{\star}\right> \mathrm{Im\;}\left< \bq,\left(\ba_0 \right)_{\star}\right>-\mathrm{Re\;}\left<\bq,\left(\ba_0 \right)_{\star}\right> \mathrm{Im\;}\left< \bq,N \cdot \left(\ba_0 \right)_{\star}\right>}{\left|\left< \bq,N \cdot \left(\ba_0 \right)_{\star}\right> \right|^2} \;.
\label{imtstar}
\ee

The central charge evaluated on the attractor locus is 
\be
\left|Z\left(\bq\right)\right|^2_{\star} = \frac{2\;\mathrm{Im}\left(\left<\bq,\left(\ba_0 \right)_{\star}\right>\left<\bq,N\cdot\left(\overline{\ba}_0 \right)_{\star}\right> \right)}{\left<\left(\ba_0 \right)_{\star},N\cdot \left(\overline{\ba}_0 \right)_{\star} \right>} \;.
\label{entd1}
\ee
If the conditions (\ref{attrimgt}) and (\ref{zisphys}) are satisfied, so the attractor locus is physical, then this gives the entropy of the single-centered black hole solution and so the associated contribution to the spectrum of BPS states of that charge. It is worth noting that the central charge on the attractor locus (\ref{entd1}) is invariant under the monodromy transformation (\ref{monchmap}). This is a general result, holding not only in this specific case.

The existence of a nilpotent orbit expansion implies that there is an associated splitting of the charges into electric and magnetic as in (\ref{n1d1dyelma}). A charge which admits a controlled and physical attractor locus, so satisfying (\ref{attrimgt}) and (\ref{zisphys}), must be magnetic. This is clear from the expression of the central charge (\ref{Ztd1n1}): purely electric charges only depend on the moduli through an overall factor which cannot lead to a solution to the attractor equations. 

In general, it is not difficult to find infinite classes of charges which have controlled attractor loci, so satisfy (\ref{attrimgt}) and (\ref{zisphys}). Let us consider such a charge and denote it $\bq^{\mathrm{Phys}}$. It can have a magnetic and an electric component, 
%Further, since we require some controlled large value of $t$ for (\ref{attrimgt}) we can expect that it must have some large charges. We will see that those charges must be electric in type. 
so we can write 
\be
\bq^{\mathrm{Phys}} = \bq_m + Q \bq_e \;.
\label{qphyformass}
\ee
Here $\bq_m$ is the magnetic component of the charge, $\bq_e$ is the electric component, and $Q$ is some integer. We can then consider, schematically, 
\be
\bq_m \sim {\cal O}(1) \;,\;\; \bq_e \sim {\cal O}(1) \;,\;\; \left|Q\right| \gg 1 \;. 
\label{proc1l}
\ee
%We take $Q \gg 1$.\footnote{More generally we should take a sum over different electric charges, so $\sum_i Q_i\bq_e^i$. However, this will not modify any of the analysis and so it is sufficient to consider the simplest case.} 
%We can then write schematically
%\be
%\bq_d \sim {\cal O}(1) \;,\;\; \bq_e \sim {\cal O}(1) \;,\;\;  Q \gg 1 \;.
%\label{qphysapp}
%\ee
%We expect that the form (\ref{qphyformass}-\ref{qphysapp}) is general for attractor values satisfying (\ref{attrimgt}) and (\ref{zisphys}).  
Substituting the form (\ref{qphyformass}) into (\ref{tstb}), and using the properties (\ref{n1d1dyelma}) and (\ref{proc1l}) gives 
\be
\mathrm{Im\;}t_{\star} \simeq Q\;\mathrm{Im} \left( \frac{\left<\bq_e,\left(\ba_0 \right)_{\star}\right>}{\left<\bq_m,N \cdot \left(\ba_0 \right)_{\star}\right>} \right) \;.
\label{attlargeQ}
\ee
We see that by choosing $Q$ large and of appropriate sign we can always satisfy (\ref{attrimgt}). 
%So let us consider charges of the schematic type
%\be
%\bq_m \sim {\cal O}(1) \;,\;\; \bq_e \sim {\cal O}(1) \;,\;\;  Q \gg 1 \;.
%\label{qphysapp}
%\ee

What about the condition (\ref{zisphys})? Let us `integrate out' the field $t$ by using its attractor locus (\ref{tstb}), but keeping $\ba_0$ as a function of the $z^i$. We reach
%We further take the behaviour (\ref{qphysapp}), and so we have
\be
Z\left(\bq,t_{\star},z^i\right) \simeq \frac{Q}{\left| c\left(z^i\right)\mathrm{Im\;}t_{\star} \right|^{\frac{1}{2}}} \left[ \left<\bq_e,\ba_0\left(z^i\right)\right>  - \left<\bq_m,N\cdot\ba_0\left(z^i\right)\right> \frac{\left<\bq_e,\overline{\ba_0\left(z^i\right)}\right>}{\left<\bq_m,N \cdot \overline{\ba_0\left(z^i\right)} \right>} + ... \right] \;. 
\label{Ztd1n1int}
\ee
This expression gives us the (approximate) form of the potential for the $z^i$, and the condition (\ref{zisphys}) is the assumption that there is a minimum to this potential in the bulk of the moduli space. An important point is that $Q$ factors out in (\ref{Ztd1n1int}) which means that the attractor locus for the $z^i$ is independent of $Q$. Since it is determined only by the choices of $\bq_m$ and $\bq_e$, which are completely general, it is in general a rather weak assumption that for some choice of charges there is a physical attractor locus satisfying (\ref{zisphys}). 

In conclusion, we have for charges of the form (\ref{qphyformass}), we expect to be able to find a physical attractor locus given the freedom to choose $Q$.

%We are interested in whether those solutions survive a small (relative to $Q$) perturbation in the electric charge. So we consider 
%\be
%\bq^{\mathrm{\delta}} = \bq_m + Q \bq_e + \delta \bq_e \;,
%\label{qphydelta}
%\ee
%where $q_m$ and $q_e$ are as in (\ref{qphyformass}), so chosen to lead to a physical attractor value, and $\delta q_e$ is electric type charge. It is a small perturbation in the sense that we take
%\be
%\frac{1}{Q}\frac{\left<\delta\bq_e,\ba_0\right>}{\left<\bq_e,\ba_0\right>} \ll 1 \;.
%\label{elpertQ}
%\ee
%The perturbation $\delta q_e$ only gives $Q$-suppressed corrections to the attractor loci determined by (\ref{attlargeQ}) and (\ref{Ztd1n1int}). Therefore, as long as (\ref{elpertQ}) is satisfied, any physical attractor loci will stay physical under the perturbation $\delta q_e$. 

%%%%%%%%%%%%%%%%%%%%
\section{Evidence for obstruction to split attractor flows}
\label{sec:split}
%%%%%%%%%%%%%%%%%%%%

In this section we study multi-centre, so split, attractor flows. These are solutions where the attractor flow crosses a wall of marginal stability and splits on it. Such flows combine the analysis of single-centred attractor flows in section \ref{sec:attflnd1}, and the analysis of walls of marginal stability in section \ref{sec:wmsmon}. Our analysis will be restricted first to single-parameter moduli spaces, and to infinite distance loci of type $d=1$. In such cases, $\ba_0$ is a true constant, which makes things much simpler to analyse. We will then go on to discuss increasingly general cases building on this. 

While we analyse things completely generally here, it is always useful to be guided by an example. In appendix \ref{sec:oneparamexam} we present such an example.

We begin with the fact that if we are free to choose the charges $\bq$, then this translates into freedom in choosing $\left<\bq,\ba_0 \right>$ and $\left<\bq,N\cdot\ba_0 \right>$. This follows from the fact that minimally there are two real charge components of $\bq$ which appear in $\left<\bq,\ba_0 \right>$, that allow us to choose its real and imaginary parts arbitrarily. This is true up to quantization of the charges, but of course if we there is an obstruction to split flows allowing for continuous charges, then it will hold for quantized charges.\footnote{In any case, it turns out that quantization of the charges does not place strong restrictions as all the relevant physical quantities have rational coefficients.} More precisely, $\left<\bq,\ba_0 \right>$ contains in general four real parameters (exactly four for one-parameter models), corresponding to two electric and two magnetic charges. It is possible to go to a basis, which we will not utilise, where it contains only two electric charges. The expression $\left<\bq,N\cdot\ba_0 \right>$ contains at least two real magnetic charges. Choosing these charges is mapped to choosing the real and imaginary parts of $\left<\bq,N\cdot\ba_0 \right>$.

We will sometimes encounter rather lengthy expressions, and so it is useful to introduce some condensed notation:
\bea
\mathrm{Re}\left<\bq^A,\ba_0 \right> &\equiv& q^{A,+} \;,\;\; \mathrm{Im}\left<\bq^A,\ba_0 \right> \equiv q^{A,-} \;, \nn \\
\mathrm{Re}\left<\bq^A,N\cdot\ba_0 \right> &\equiv& q_N^{A,+} \;,\;\; \mathrm{Im}\left<\bq^A,N\cdot\ba_0 \right> \equiv q_N^{A,-} \;.
\label{qsdef}
\eea
Since, as explained above, we are free to choose $q^{\pm}$ and $q_N^{\pm}$ arbitrarily as real parameters, the notation should hopefully not lead to any loss of clarity. Also, we recall, that since we are in one-parameter models, the $q$'s are all true constants.

%%%%%%%%%%%%%%%%%%%%
\subsection{Obstruction for one-parameter two-centre electric flows}
\label{sec:nogo2elec}
%%%%%%%%%%%%%%%%%%%%

Let us consider a setup where a purely electric charge flow splits into two dyonic flows.  We would like to know if the dyonic flows can end on physical attractor loci. We take the split as $\bq^A=\bq^B+\bq^C$, and take the magnetic component of  $\bq^A$ to vanish, $\bq^A_m=0$. 

We consider the specification of the wall of marginal stability given by (\ref{wmsZc1}) and (\ref{wmsZc2}). We can solve (\ref{wmsZc1}) for $\mathrm{Re\;}t_{WMS}$ as
\be
\mathrm{Re\;}t_{WMS} = \frac{-q^{A,-}q^{B,+} + q^{B,-}q^{A,+} + \left( q^{A,-}q_N^{B,-}   + q^{A,+}q_N^{B,+} \right) \mathrm{Im\;}t_{WMS}}{- q_N^{B,-}q^{A,+} + q^{A,-}q_N^{B,+}}\;.
\label{rtwssole2} 
\ee
Since $\mathrm{Re\;}t_{WMS}$ has no physical constraints on its sign or magnitude, there is always such a solution. Note that we are here allowing $\mathrm{Re\;}t$ to be arbitrary, while it is often considered to lie in the range $0 \leq \mathrm{Re\;}t \leq 1$. It can always be brought into this range by acting with the monodromy (gauge) transformation $\mathrm{Re\;}t \rightarrow \mathrm{Re\;}t + n$ and $\bq \rightarrow T^{n} \cdot \bq$. In other words, there is a choice of charge representative within each monodromy orbit for which $\left|\mathrm{Re\;}t\right| \leq 1$, but since we do not want to restrict the charge to that representative, we should equally keep $\mathrm{Re\;}t$ arbitrary. 

We still have the condition (\ref{wmsZc2}) to impose. First though, let us solve for the attractor loci $\mathrm{Im\;}t_{\star}^B$ and  $\mathrm{Im\;}t_{\star}^C$  of the two dyonic states $\bq^B$ and $\bq^C$, as given by (\ref{imtstar}). We do this by keeping $\mathrm{Im\;}t_{\star}^B$ and  $\mathrm{Im\;}t_{\star}^C$ arbitrary, and solving for some of the flux parameters, or equivalently the $q$'s, in terms of them as
\bea
q^{B,-} &=& \frac{q_N^{B,-} q^{B,+} + \left|q_N^{B}\right|^2 \mathrm{Im\;}t_{\star}^B }{q_N^{B,+}} \;, \nn \\
q^{A,-} &=& \frac{q_N^{B,-} q^{A,+} + \left|q_N^{B}\right|^2 \left(\mathrm{Im\;}t_{\star}^B - \mathrm{Im\;}t_{\star}^C\right)}{q_N^{B,+}} \;,
\label{attloc2ebc}
\eea
where $\left|q_N^{B}\right|^2= \left(q_N^{B,+}\right)^2+\left(q_N^{B,-}\right)^2$.

Now we can return to the second constraint for the wall of marginal stability (\ref{wmsZc2}). Substituting (\ref{rtwssole2}) and (\ref{attloc2ebc}) into (\ref{wmsZc2}) gives
\be
\frac{\mathrm{Im\;}t^B_{\star}-\mathrm{Im\;}t^C_{\star}}{ 
\mathrm{Im\;}t^B_{\star} +\mathrm{Im\;}t_{WMS}} \geq 1\;. 
\label{nogo2c}
\ee
It is manifest now that there is no way to satisfy this inequality while maintaining the physical constraint on both the attractor loci as well as the location of the wall of marginal stability 
\be
\mathrm{Im\;}t^B_{\star}  > 0 \;,\;\; \mathrm{Im\;}t^C_{\star}  > 0 \;,\;\; \mathrm{Im\;}t_{WMS} > 0 \;.
\ee
We have therefore proven that, in the one-parameter moduli space case, that there are no two-centre physical split attractor flows with total electric charge, at least not in the weakly-coupled region of moduli space $\mathrm{Im\;}t \gg 1$. 

There is no obstruction to a split flow which has a negative attractor locus for one of the dyons. We can take, for example, $-\mathrm{Im\;}t^C_{\star} > \mathrm{Im\;}t_{WMS} > 0$ and satisfy the condition (\ref{nogo2c}). What does it mean that we have an attractor locus which has $\mathrm{Im\;}t_{\star} < 0$? This just means that the attractor flow will drive $ \mathrm{Im\;}t \rightarrow 1$  approaching the black hole horizon. What happens after that is not clear because by then we are in the strongly-coupled regime of the moduli space, where the exponentially suppressed (instanton-type) corrections in $e^{-\mathrm{Im\;}t}$ become important and the Nilpotent orbit approximation breaks down.

It is worthwhile looking at why the obstruction comes about in a less sharp but more direct way. The requirement that the total charge is purely electric imposes that the magnetic charges of the two dyons are opposite $\bq_m^B = - \bq_m^C$. From (\ref{imtstar}) we therefore see that if $\mathrm{Im\;}\left<\bq_e^B,\left(\ba_0\right)_{\star} \right>$ and $\mathrm{Im\;}\left<\bq_e^C,\left(\ba_0\right)_{\star} \right>$ have the same sign (and also the real parts), then we would have that $\mathrm{Im\;}t^B_{\star}$ and $\mathrm{Im\;}t^C_{\star}$ must have opposite signs and so they cannot both be positive.\footnote{The magnetic parts of the charges in $\left<\bq^B,\left(\ba_0\right)_{\star} \right>$ cancel out in (\ref{imtstar}).} We also have that the sum of the electric charges sum to the total electric charge $\bq^A_e = \bq^B_e + \bq^C_e$. But we need that the electric state is more massive than the two dyonic states, and it obtains all its mass from the electric charge. So we expect, schematically, $\left|\bq^A_e\right|  > \left|\bq^B_e\right|$ and $\left|\bq^A_e\right|  > \left|\bq^C_e\right|$. So the electric charges $\bq^B_e$ and $\bq^C_e$ must `add up' rather than `subtract', and this results in  $\mathrm{Im\;}\left<\bq_e^B,\left(\ba_0\right)_{\star} \right>$ and $\mathrm{Im\;}\left<\bq_e^C,\left(\ba_0\right)_{\star} \right>$ having the same sign. This picture is of course more schematic than the sharp result (\ref{nogo2c}), but it makes manifest the underlying cause.

What can we make of this no-go in terms of the stability of the electric states? If the attractor flows ended in well-controlled loci, we would be able to deduce that the dyonic constituents were in the spectrum of BPS states. Therefore, crossing the wall of marginal stability in the opposite direction, also the electric state would be in the spectrum. Since it would be heavier than the dyonic constituents, which are themselves heavier than the Planck scale, it would be a black hole. This black hole would then be unstable and decay to the dyons upon the path to weak coupling which would make it sub-Planckian in mass eventually. Also, following the discussion in section \ref{sec:BPSCY}, we would expect that by choosing the electric charge to be of the form $\bq^A_e = K \bq_e$, for some large integer $K$, we can deduce the existence of an electric state of charge $\bq_e$ in the spectrum. But this state would be sub-Planckian and cannot decay to dyonic states upon the path to weak coupling. It would therefore remain stable, and this way we could populate the BPS states of the light tower.

If the no-go would imply that there are just no physical attractor flows for both the dyons, then we would deduce that $\bq^B$ and $\bq^C$ cannot both be populated by BPS states at the same time. This means that if the black hole $\bq^A$ was in the spectrum, so is physical, then it will remain stable upon the path to weak coupling and become part of the sub-Planckian light tower. 

The fact that the attractor flows must go into the strongly-coupled region, means at this general level, the results remain ambiguous. 

%So far we have not restricted in any way the choice of charges. In particular, $\bq^B$ is arbitrary which means that, as discussed in section \ref{sec:attflnd1}, we can always choose it such that it leads to a physical attractor locus. So given an attractor flow which starts off purely electric at infinity, after crossing a wall of marginal stability it can split into two dyonic flows, and at least one of these dyonic flows can be made to terminate on a physical attractor locus. The non-trivial question is if both of the dyonic attractor loci can be made physical (positive values of $\mathrm{Im\;}t_{\star}$). 

%%%%%%%%%%%%%%%%%%%%
\subsection{Obstruction for one-parameter $n$-centre electric flows}
\label{sec:nogoNelec}
%%%%%%%%%%%%%%%%%%%%

After the first split of the attractors, one of the dyonic flows is towards small $\mathrm{Im\;t}$. But we can consider the flow splitting again, this time into two dyons with well controlled attractor loci. This can occur an arbitrary number of times, and so we should consider in generality an $n$-centre black hole configuration, where each centre is a dyonic black hole with a controlled attractor locus, and the total sum of the charges is purely electric. 

To show that this is not possible, we will show the following result: a dyonic flow towards a negative attractor value for $\mathrm{Im\;}t_{\star}$ cannot split into two dyonic flows into positive attractor values. So this means that there is no way to get rid of the negative attractor value we found from the two-centre solution, so the one induced after the first split, no matter how many subsequent splits there are. 

This is a little more involved to show than the result of section \ref{sec:nogo2elec}, but follows similar logic. We will eliminate fluxes for the attractor and wall of marginal stability values of the modulus, and show an inconsistency. So we consider $\bq^A=\bq^B+\bq^C$, but now all the charges are dyonic. The only constraint we will impose is that $\bq^A$ has a negative value for its attractor locus
\be
\mathrm{Im\;} t^A_{\star} < 0 \;.
\ee
We can eliminate some fluxes for the attractor loci of B and C by solving (\ref{imtstar}) to give
\be
q^{B,-} = \frac{q_N^{B,-} q^{B,+} + \left|q_N^{B}\right|^2 \mathrm{Im\;}t_{\star}^B }{q_N^{B,+}} \;, \;\;
q^{C,-} = \frac{q_N^{C,-} q^{C,+} + \left|q_N^{C}\right|^2 \mathrm{Im\;}t_{\star}^C }{q_N^{C,+}} \;.
\label{attlocnbc}
\ee
We can then substitute (\ref{attlocnbc}) into the attractor locus for state A, and solve that as
\bea
q^{B,+} = &-&\frac{1}{q_N^{C,+} \left(-q_N^{C,-} q_N^{B,+} + q_N^{B,-} q_N^{C,+}\right)} \label{attsome} \\
& & \Bigg[ q_N^{C,-} q_N^{B,+} \left(q^{C,+} q_N^{B,+} - 2 q_N^{B,-} q_N^{C,+} \mathrm{Im\;}t_{\star}^A\right) \nn \\
& & + \left(q_N^{C,-}\right)^2 q_N^{B,+} \left(q_N^{B,+} \mathrm{Im\;}t_{\star}^C + q_N^{C,+} \left(\mathrm{Im\;}t_{\star}^C - \mathrm{Im\;}t_{\star}^A\right)\right) \nn \\
& & + 
     q_N^{C,+} \left(-q_N^{B,-} q^{C,+} q_N^{B,+} + 
        \left(q_N^{B,-}\right)^2 \left(q_N^{C,+} \mathrm{Im\;}t_{\star}^B + q_N^{B,+} \left(\mathrm{Im\;}t_{\star}^B - \mathrm{Im\;}t_{\star}^A\right)\right)\right) \nn \\
        & & + 
        q_N^{B,+} \left(q_N^{B,+} + q_N^{C,+}\right) \left(q_N^{B,+} \left(\mathrm{Im\;}t_{\star}^B - \mathrm{Im\;}t_{\star}^A\right) + 
           q_N^{C,+} \left(\mathrm{Im\;}t_{\star}^C - \mathrm{Im\;}t_{\star}^A\right)\right)\Bigg] \nn
\eea

For the wall of marginal stability we will take the conditions (\ref{bcwmsZc1}) and (\ref{bcwmsZc2}). We again would like to solve the constraint (\ref{bcwmsZc1}) for $\mathrm{Re\;} t_{WMS}$. However, unlike the purely electric case, the constraint (\ref{bcwmsZc1})  is now quadratic in  $\mathrm{Re\;} t_{WMS}$ which means that a real solution is not guaranteed. We should require some positivity constraint on the coefficients. More precisely, we can solve (\ref{bcwmsZc1}) as
\bea
\mathrm{Re\;} t_{WMS} = \frac{-q_N^{C,-} q^{B,+} + q_N^{B,-} q^{C,+}  - q^{C,-} q^{B,+} + q_N^{C,+} q^{B,-} -\sqrt{P}}{2\left( q_N^{C,-} q_N^{B,+} -  q_N^{B,-} q_N^{C,+} \right)} \;,
\eea
where $P$ is a quantity which we require to be positive to obtain a real solution
\bea
P &=& \left(q_N^{C,-} q^{B,+} - q_N^{B,-} q^{C,+} + q^{C,-} q_N^{B,+} - q^{B,-} q_N^{C,+}\right)^2  \nn \\
& &- 4 \left( q_N^{C,-} q_N^{B,+} - q_N^{B,-} q_N^{C,+} \right) \Bigg[-q^{B,-} q^{C,+} + q^{C,-} \left(q^{B,+} - q_N^{B,-} \;\mathrm{Im\;}t_{WMS}\right) \nn \\
& &+ 
    \mathrm{Im\;}t_{WMS} \left(q^{B,-} q_N^{C,-} - q^{C,+} q_N^{B,+} + q^{B,+} q_N^{C,+} \right. \nn \\
    & & \left.+ q_N^{C,-} q_N^{B,+}\; \mathrm{Im\;}t_{WMS} - 
       q_N^{B,-} q_N^{C,+} \;\mathrm{Im\;}t_{WMS}\right) \Bigg] > 0 \;. 
\label{cons110}
\eea
The constraint for a solution (\ref{cons110}) must then be imposed along with the second constraint on the wall of marginal stability (\ref{bcwmsZc2}). After substituting the attractor values (\ref{attlocnbc}) and (\ref{attsome}), the two constraints can be written as
\be
P = X^2 - Y > 0 \;,\;\; X + \sqrt{X^2 - Y} > 0 \;,
\label{fnconsts} 
\ee
where
\bea
X &=&  2 \left(q^{B,-}_N q^{C,-}_N +  q^{B,+}_N q^{C,+}_N\right) \mathrm{Im\;} t_{WMS}  \nn \\
& & + \left[\left(q^{B,-}_N + q^{C,-}_N \right)^2 + \left(q^{B,+}_N + q^{C,+}_N \right)^2  \right] \mathrm{Im\;} t_{\star}^A 
 - \left|q^{B}_N\right|^2\mathrm{Im\;} t_{\star}^B  - \left|q^{C}_N\right|^2 \mathrm{Im\;} t_{\star}^C \;, \label{Xdef}  \nn \\
Y &=& 4 \left|q^{B}_N\right|^2\left|q^{C}_N\right|^2  (\mathrm{Im\;} t_{WMS} + \mathrm{Im\;} t_{\star}^B) (\mathrm{Im\;} t_{WMS} + \mathrm{Im\;} t_{\star}^C)
\label{Ydef}
\eea
A solution to the constraints (\ref{fnconsts}) requires $X > 0$. Now since we are interested in attractor loci of the form
\be
\mathrm{Im\;} t_{\star}^A < 0 \;,\;\; \mathrm{Im\;} t_{\star}^B > 0 \;,\;\; \mathrm{Im\;} t_{\star}^C > 0 \;, 
\label{attloccho}
\ee
we see that 
\be
 2 \left(q^{B,-}_N q^{C,-}_N +  q^{B,+}_N q^{C,+}_N\right) \mathrm{Im\;} t_{WMS} > X > 0 \;.
\ee
But (\ref{fnconsts}) also requires $X^2 > Y$, so we require
\be
\left(2 \left(q^{B,-}_N q^{C,-}_N +  q^{B,+}_N q^{C,+}_N\right) \mathrm{Im\;} t_{WMS}\right)^2 > X^2 > Y > \Big(2 \left|q^{B}_N\right|\left|q^{C}_N\right| \mathrm{Im\;} t_{WMS} \Big)^2 \;.
\label{Prefo}
\ee
It is a simple identity that 
\be
\left(q^{B,-}_N q^{C,-}_N +  q^{B,+}_N q^{C,+}_N\right)^2 < \left|q^{B}_N\right|^2\left|q^{C}_N\right|^2 \;,
\ee
and therefore (\ref{Prefo}) cannot be satisfied. Hence, it is not possible to satisfy the constraints for the wall of marginal stability (\ref{fnconsts}) and the attractor loci (\ref{attloccho}) simultaneously. This proves that any attractor flow towards a negative attractor locus, cannot split into two attractor flows which are both towards positive attractor loci. 

%%%%%%%%%%%%%%%%%%%%
\subsection{Generalizing to multi-parameter models}
\label{sec:nogogen}
%%%%%%%%%%%%%%%%%%%%

We have shown that for one-parameter moduli spaces, there are no split attractor flows near $n=d=1$ infinite distance loci which have a total electric charge and terminate on controlled attractor loci. In this section we generalise this result. 

Generalising to the case of an arbitrary number of moduli still allows the natural restriction of one-parameter approaches to infinite distances, and so there is a singled-out coordinate $t$ which controls the infinite distance limit. The crucial difference from the analysis of sections \ref{sec:nogo2elec} and \ref{sec:nogoNelec} is that now the limiting period vector is not a constant but a function of the other moduli $\ba_0 = \ba_0\left(z_i\right)$. This first of all places additional constraints on the attractor loci, that they should be at controlled and physical values of the $z_i$. However, this is expected to be rather mild, and in any case can only make the constraints on the existence of split flows stronger. 

The aspect which may seem to allow to evade the no-go results, at least on first thought, is that the $z_i$ moduli can vary between the locus of marginal stability and the attractor loci. So $\left. \ba_0\left(z_i\right) \right|_{WMS} $ and $\ba_0\left(z_i\right)_{\star}$ need not be the same which would imply that the $q$'s in (\ref{qsdef}) also need not be the same. Recall that it was the interaction between these two loci in moduli space which led to the obstruction of the split attractor flows. The wall of marginal stability and the attractor loci could individually be made positive (physical), but not all simultaneously. In this section we will study in examples if this additional freedom allows to avoid the obstruction. 

%%%%%%%%%%%%%%%%%%%%
\subsubsection{Obstruction in a two-parameter example}
\label{sec:nogotwoNelec}
%%%%%%%%%%%%%%%%%%%%

In this section we study the case of a two-parameter model near the large complex-structure point. We will consider the mirror to the Calabi-Yau $\mathbb{P}_4^{(1,1,2,2,2)}[8]$ as studied in \cite{Candelas:1993dm}. There are two monodromies associated to the two complex moduli fields $t_1$ and $t_2$. These take the form
\be
N_1 = \left( \begin{array}{cccccc} 
0 & 0 & 0 & 0 & 0 & 0\\ 
1 & 0 & 0 & 0 & 0 & 0\\ 
0 & 0 & 0 & 0 & 0 & 0\\ 
-2 & -4 & 0 & 0 & 0 & 0\\ 
0 & -8 & -4 & 0 & 0 & 0\\ 
-\frac{22}{3} & 0 & -2 & 0 & -1 & 0
\end{array} \right) \;, \;\; 
N_2 = \left( \begin{array}{cccccc} 
0 & 0 & 0 & 0 & 0 & 0\\ 
0 & 0 & 0 & 0 & 0 & 0\\ 
1 & 0 & 0 & 0 & 0 & 0\\ 
0 & 0 & 0 & 0 & 0 & 0\\ 
-2 & -4 & 0 & 0 & 0 & 0\\ 
-2 & -2 & 0 & -1 & 0 & 0
\end{array} \right) \;. 
\ee
We note that the nilpotency indices are $n_1=3$ and $n_2=1$, so $N_1^4=0$ and $N_2^2=0$. We can write the Nilpotent orbit of the period vector as
\be
\Pi\left(t_1,t_2\right) = e^{N_1 t_1} \ba^{(1)}_0\left(t_2\right) = e^{N_2 t_2} \ba^{(2)}_{0}\left(t_1\right) = e^{N_2t_2 + N_1t_1} \ba_0 \;,
\ee
where the limiting vectors are
\be
\ba^{(1)}_0\left(t_2\right) = \left(\begin{array}{c} 1 \\ 0 \\ t_2 \\ -1  \\ -\frac{11}{3}-2t_2 \\ -t_2+2 i \xi \end{array}\right) \;,\;\;
\ba^{(2)}_0\left(t_1\right) = \left(\begin{array}{c} 1 \\ t_1 \\ 0 \\ -1 -2t_1 -2t_1^2 \\ -\frac{11}{3}-4t_1^2 \\ \frac13\left( -11t_1+4t_1^2+6 i \xi\right) \end{array}\right) \;,\;\;
\ba_0 = \left(\begin{array}{c} 1 \\ 0 \\ 0 \\ -1 \\ -\frac{11}{3}\\ 2 i \xi\end{array}\right) \;,
\ee
where $\xi$ is an order-one real constant whose value is not important for our purposes. 
Note that in this example we have taken the symplectic inner-product matrix as
\be
\eta = \left( \begin{array}{cccccc} 
0 & 0 & 0 & 0 & 0 & -1\\ 
0 & 0 & 0 & 0 & -1 & 0\\ 
0 & 0 & 0 & -1 & 0 & 0\\ 
0 & 0 & 1 & 0 & 0 & 0\\ 
0 & 1 & 0 & 0 & 0 & 0\\ 
1 & 0 & 0 & 0 & 0 & 0
\end{array} \right) \;, 
\ee
which is slightly different to (\ref{etabas}).

Denoting the real and imaginary parts of the $t_i$ as
\be
t_i = b_i + iv_i \;,
\ee
there are two types of infinite distance limits we can consider which match onto the formulation in this paper:
\bea
n=d=1\;&:&\;  v_1 \mathrm{\;finite}\;,\;\; v_2 \rightarrow \infty \;, \nn \\
n=d=3\;&:&\;  v_1 \rightarrow \infty \;,\;\; v_2 \mathrm{\;finite} \;.
\eea
Note that to keep control of the nilpotent orbit expressions we need to keep always $v_1,v_2 \gg 1$. 

\subsubsection*{Simplifying assumption}

Let us denote the general charge vector as
\be
\bq = \left(q_0,q_1,q_2,q_3,q_4,q_5\right)^T \;.
\ee
It is difficult to analyse the attractor loci for charges in full generality. We will therefore simplify the analysis by restricting to the case where all charges have  
\be
q_0=0 \;.
\label{assq00}
\ee
It is important to keep in mind that the obstruction we find may be related to this restriction, and there exists the possibility that relaxing it will allow a realisation of the split attractor flows. But in any case this is an example analysis, not a general one, and it aim is to test for an obstruction once we have multi-parameters, as relevant for the discussion in section \ref{sec:nogogen}. At least this aspect is not affected by the restriction (\ref{assq00}).

\subsubsection{The attractor loci}

We would like to solve for the attractor loci, specifically in order to test if they are at physical well-controlled points in moduli space. Keeping in mind this aim of the analysis, we can make a further simplification
\be
\xi \rightarrow 0 \;.
\ee
The justification for this is that because of the restriction (\ref{assq00}), the only place where $\xi$ appears is in the Kahler potential for the moduli. As long as $v_1,v_2 \gg 1$, it gives a small correction which cannot change the validity of an attractor locus. 

With the states simplifications, the attractor loci are
\bea
\left(v_1\right)_{\star} &=& \frac12\sqrt{Q_{\star}} \;,\;\;
\left(v_2\right)_{\star} = \frac{q_2}{q_1}\left(v_1\right)_{\star} \;, \nn \\
\left(b_1\right)_{\star} &=& -\frac{2 q_1 + q_3}{4q_1} \;,\;\;
\left(b_2\right)_{\star} = \frac{4q_1^2 +2 q_1 q_3+q_2q_3-q_1 q_4}{4q_1^2}\;,
\label{attsol2px1}
\eea
where
\be
Q_{\star} = \frac{-112 q_1^3 - 3 q_2 q_3^2 - 6 q_1 q_3 \left(q_3 - q_4\right) - 12 q_1^2 \left(q_2 + 2 q_3 - q_4 + 2 q_5\right)}{q_1^2\left(2q_1+3q_2\right)} \;.
 \label{attsol2px2}
\ee
Here we have assumed $q_1 \neq 0$ which, as we will see, guarantees that the state is dyonic with respect to the electromagnetic splitting induced by both $N_1$ and $N_2$. Recall that being dyonic is a necessary requirement for a non-divergent attractor locus. 

\subsubsection{The $n=d=1$ sector}
\label{sec:nd1at}

Let us consider first the case of $v_2 \rightarrow \infty$ as the degeneration limit, so it is an $n=d=1$ degeneration. First we can split the charges into electric and magnetic as in (\ref{n1d1dyelma}). We have 
\be
\bq_e = \left( \begin{array}{c} 0 \\ 0 \\ q_2 \\ 0 \\ q_4 \\ q_5 \end{array}\right) \;,\;\; 
\bq_m = \left( \begin{array}{c} q_0 \\ q_1 \\ 0 \\ q_3 \\ 0 \\ 0 \end{array}\right) \;,\;\; 
\ee
We are interested in split attractor flows of the type $\bq^A_e \rightarrow \bq^B_d + \bq^C_d$, and define the wall of marginal stability by the constraints (\ref{bcwmsZc1}) and (\ref{bcwmsZc2}). We can solve (\ref{bcwmsZc1}) for $b_2$ as
\bea
b_2 &=& \frac{1}{3 v_1 \left(8 b_1^2 q^A_2 q^B_1 - 4 q^A_5 q^B_1 + 
     2 q^A_2 q^B_3 + 4 b_1 q^A_2 \left(2 q^B_1 + q^B_3\right) + q^A_4 (2 q^B_1 + q^B_3) + 
     8 q^A_2 q^B_1 v_1^2\right)} \times\nn \\
& &\Bigg[3 q^A_5 \left(2 q^B_2 + 4 b_1 \left(2 q^B_1 + q^B_2\right) + q^B_4\right) v_1 + 
   q^A_4 v_1 \left[-11 q^B_1 + 12 b_1^2 q^B_1 + 6 b_1^2 q^B_2 - 3 \left(q^B_2 + q^B_5\right) \right.\nn \\
   & & \left.+ 6 \left(2 q^B_1 + q^B_2\right) v_1^2 \right] + 
   q^A_2 v_1 \left[q^B_1 \left(-22 + 4 b_1 \left(-5 + 6 b_1\right) + 24 v_1^2\right) \right.\nn \\
  & & \left.- 3 \left(2 q^B_5 + 4 b_1 q^B_5 + q^B_4 \left(-1 + 2 b_1^2 + 2 v_1^2\right)\right)\right] \nn \\
  & & + 
   3 \left(q^A_2 + 2 b_1 \left(1 + b_1\right) q^A_2 + b_1 q^A_4 + q^A_5\right) \left(\left(2 + 4 b_1\right) q^B_1 + 
      q^B_3\right) v_2 \nn \\
      & & + 6 \left(2 \left(q^A_2 + 2 b_1 q^A_2 + q^A_4\right) q^B_1 - 
      q^A_2 q^B_3\right) v_1^2 v_2 \Bigg] \;.
\label{solwmscns1b2}
\eea
Substituting this solution into (\ref{bcwmsZc2}), gives the constraint we need to satisfy
\bea
& &R_{AB} \equiv \Bigg(3 v_1 \left(8 b_1^2 q^A_2 q^B_1 - 4 q^A_5 q^B_1 + 2 q^A_2 q^B_3 + 
     4 b_1 q^A_2 \left(2 q^B_1 + q^B_3\right) + q^A_4 \left(2 q^B_1 + q^B_3\right) + 
     8 q^A_2 q^B_1 v_1^2\right)\Bigg) \nn \\
     & & \Bigg[v_1 \left(3 q^B_3 \left(\left(2 + 4 b_1\right) q^B_2 + q^B_4\right) + 
      \left(q^B_1\right)^2 \left(-44 + 48 b_1 \left(1 + b_1\right) + 48 v_1^2\right) \right. \nn \\
     & & \left.+ 
      6 q^B_1 \left(4 b_1 \left(q^B_2 + b_1 q^B_2 + q^B_3\right) + q^B_4 - 2 q^B_5 + 4 q^B_2 v_1^2\right)\right) \nn \\
      & &+ 
   3 \left(\left(\left(2 + 4 b_1\right) q^B_1 + q^B_3\right)^2 + 16 \left(q^B_1\right)^2 v_1^2\right) v_2\Bigg]^{-1} \geq 1 \;.
\label{cons22pb2}
\eea

At this point we need to impose that the attractor values for states $B$ and $C$ are physical. The attractor solutions are given by (\ref{attsol2px1}-\ref{attsol2px2}). We can eliminate generally $q^B_5$ and $q^A_5$ for $Q_{\star}^B$ and $Q_{\star}^C$. Once we do that, the physical attractor loci conditions are
\be
Q_{\star}^B > 0 \;,\;\; Q_{\star}^C > 0 \;,\;\; \frac{q_2^B}{q^B_1} > 0 \;,\;\; \frac{q_2^B-q^A_2}{q^B_1} > 0 \;.
\label{pysatcon2p}
\ee

Now let us return to the constraint (\ref{cons22pb2}). We note that that we have a free real unconstrained parameter $b_1$ in the constraint. We can consider eliminating $b_1$ for $R_{AB}$ (as defined in (\ref{cons22pb2})).  However, because $b_1$ appears quadratically, there is a resulting condition on the positivity of the square root to obtain a real solution (much like the constraint (\ref{cons110})). After substituting the elimination of $q^B_5$ and $q^A_5$ for $Q_{\star}^B$ and $Q_{\star}^C$ into this expression, we recover the constraint
\bea
& &-\Bigg[\left(q^B_2-q^A_2\right) v_1 + q^B_2 v_1 \left(-1 + R_{AB}\right) + 
   2 q^B_1 \left(v_1 + v_2\right) R_{AB}\Bigg]\times \nn \\
   & & \Bigg[Q_{\star}^C \left(3 \left(q^B_2-q^A_2\right) + 2 q^B_1\right) + 
   Q_{\star}^B \left(2 q^B_1 + 3 q^B_2\right) \left(-1 + R_{AB}\right) \nn \\
   & &+ 3 v_1 \left(\left(q^B_2-q^A_2\right) v_1 + q^B_2 v_1 \left(-1 + R_{AB}\right) + 2 q^B_1 \left(v_1 + v_2\right) R_{AB}\right)\Bigg] > 0 \;.
    \label{consfin2pex}
\eea
It is manifest that this constraint is incompatible with (\ref{cons22pb2}), which implies $R_{AB} \geq 1$, and with (\ref{pysatcon2p}). To see this note that (\ref{pysatcon2p}) implies that $q^B_1$, $q^B_2$ and $q^B_2-q^A_2$ must all have the same sign, and therefore all the terms in the two square brackets have the same signs. 

We have therefore shown that, up to the restriction (\ref{assq00}), there are no electric split attractor flows in this example. So again we find an obstruction. 

\subsubsection{The $n=d=3$ sector}

We can repeat the same analysis but now with the electric-magnetic splitting dictated by $N_1$. This corresponds to the limit $v_1 \rightarrow \infty$ with $v_2$ finite. In this case the electric-magnetic splitting is
\be
\bq_e = \left( \begin{array}{c} 0 \\ 0 \\ 0 \\ q_3 \\ q_4 \\ q_5 \end{array}\right) \;,\;\; 
\bq_m = \left( \begin{array}{c} q_0 \\ q_1 \\ q_2 \\ 0 \\ 0 \\ 0 \end{array}\right) \;.
\ee
The attractor loci solutions in (\ref{attsol2px1}) are still valid, but with the further restriction that, since state $A$ is electric, $q^A_2=0$. This therefore gives the constraints 
\be
Q_{\star}^B > 0 \;,\;\; Q_{\star}^C > 0 \;,\;\; \frac{q_2^B}{q^B_1} > 0 \;.
\label{pysatcon2pn1}
\ee

To determine the wall of marginal stability we can again solve for $b_2$. However, to keep things tractable in this case we make the further simplifying assumption
\be
q^A_3 =0 \;.
\ee
We can then go through the same steps as in section \ref{sec:nd1at} and we arrive at the final positivity constraint (the analogue of (\ref{consfin2pex}))
\bea
& &-R_{AB}\Bigg[q^B_2 v_1 + 2 q^B_1 \left(v_1 + v_2\right) \Bigg]\times  \\
   & & \Bigg[Q_{\star}^C \left(3 q^B_2 + 2 q^B_1\right) + 
   Q_{\star}^B \left(2 q^B_1 + 3 q^B_2\right) \left(-1 + R_{AB}\right) + 3 v_1 R_{AB} \left(q^B_2 v_1 +2 q^B_1 \left(v_1 + v_2\right) \right)\Bigg] > 0 \;. \nn 
    \label{consfin2pex}
\eea
Again there is an obstruction to (\ref{consfin2pex}) for $R_{AB} \geq 1$ and the constraints (\ref{pysatcon2pn1}).

%%%%%%%%%%%%%%%%%%%%
\section{Obstruction from moduli space singularities}
\label{sec:obspos}
%%%%%%%%%%%%%%%%%%%%

In section \ref{sec:split} we presented evidence that there are no split attractor flows where a total electric charge splits into a number of dyonic ones with well-controlled attractor loci. This suggests that electric states do not decay to dyonic states upon crossing walls of marginal stability. 

In this section we will present further strong evidence for this, and also generalise it significantly. The key idea is that in any weakly-coupled region of moduli space, where the nilpotent orbit approximation holds, we know that there are no singularities in the moduli space away from the degeneration limit $g \rightarrow 0$.\footnote{By this we mean we consider the complex plane of the modulus associated to the nilpotent orbit expansion. The other moduli can lead to singularities for non-generic values, but this is irrelevant for the flow to $g \rightarrow 0$ which can be achieved by varying the nilpotent orbit modulus.} This places constraints on the BPS spectrum, because we cannot have a populated state become massless in that region. So any charge $\bq$ for which the central charge $Z\left(\bq\right)$ vanishes in a weakly-coupled asymptotic region must be absent from the spectrum. We will show that the electric to dyonic decays are obstructed by this, so one of the dyonic constituents would always have a vanishing point along the wall of marginal stability of the decay and should therefore not be in the spectrum.

The vanishing of the central charge along walls of marginal stability is something that played an important role in the formulation of the Kontsevich-Soibelman (KS) wall-crossing formula \cite{Kontsevich:2008fj}. It was assumed that this does not occur, or more precisely the formula was restricted to states which do not have a vanishing central charge at the point where the wall is crossed. The formula applies to charges which satisfy, on the wall of marginal stability, a certain positivity constraint. Namely, there is a quadratic form on the lattice of charges, so an inner product which maps a charge to a real number ${\cal Q}\;:\; \bq \rightarrow \mathbb{R}$. Then the formula applies to {\it stable} charges, so which satisfy
\be
{\cal Q}\left(\bq\right) \geq 0 \;. 
\label{stabq}
\ee
The form further has the property that
\be
\left.{\cal Q}\right|_{\mathrm{Ker}Z\left(\bq\right)} < 0 \;.
\ee
So, charges which lead to a vanishing central charge $Z\left(\bq\right)=0$ are not stable charges.  
It was discussed in \cite{Kontsevich:2008fj} that there is a natural candidate for such a quadratic form on the Calabi-Yau. 

The stability constraint is essential because it fixes the finiteness of the number of combinations of charges which align on a wall of marginal stability. Specifically, on the wall we may consider alignment 
 \be
 \bq^A\left(r_B,r_C\right) = r_B\; \bq^B + r_C\;\bq^C \;,
 \ee
 with $r_B$ and $r_C$ some integers. So we let $\bq^A$ be a charge which depends on $r_B$ and $r_C$. Then the stability condition ensures that there is a basis of the alignment charges such that $r_B,r_C \geq 0$ (or $r_B,r_C \leq 0$) for all populated BPS states. This positivity constraint is crucial to guarantee finiteness in the wall crossing formula \cite{Kontsevich:2008fj,Andriyash:2010qv,Gaiotto:2009hg}.

Returning to the singularities on the moduli space. We see that it is the same as requiring that all the populated states on the wall of marginal stability are stable (\ref{stabq}). It is in this sense the statement that the wall-crossing formula should apply to all BPS states at any point along the wall, at least in any weakly-coupled region. 

What we are able to show, below, is that an electric to dyonic decay implies one of the dyonic states is not stable at some point on the wall. This suggests that it cannot be part of the BPS spectrum, matching the results from the attractor loci analysis. The final part of the proof is to show that it cannot be that it participated in the decay, and then itself decayed away from the spectrum before it became unstable. We can prove this for the case $n=d=1$, but have not done so completely generally for any $n$ and $d$. 
 
A key aspect of the analysis is that we find that the absence of a singularity, or the stability constraint, leads to the same expressions as the positivity of the attractor loci. This is expected if the correspondence between attractors and the spectrum of BPS states holds, and can be said to provide strong support for it. Indeed, since the attractor locus minimizes the absolute value of the central charge, we could not have a physical attractor locus with a non-vanishing central charge, and also a physical locus where the central charge vanishes. 

%%%%%%%%%%%%%%%%%%%%
\subsection{Equivalence of stability and attractor loci}
\label{sec:equonepara}
%%%%%%%%%%%%%%%%%%%%

Let us consider the one-parameter $n=d=1$ case, as studied in section \ref{sec:nogo2elec}. The wall of marginal stability is given by (\ref{rtwssole2}). We can then substitute this into the central charge formula and solve for vanishing central charge constraint on the wall
\be
\left.Z\left(\bq,\mathrm{Im\;}t_0\right) \right|_{WMS} =0 \;.
\label{vansloc}
\ee
So here we are restricting the central charge to the wall, by solving for $\mathrm{Re\;}t$ as in (\ref{rtwssole2}), and then defining $\mathrm{Im\;}t_0$ as the value of $\mathrm{Im\;}t$ where the central charge vanishes. Doing this for the two charges $\bq^B$ and $\bq^C$ gives
\bea
\mathrm{Im\;}t_0^B &=& -\left[\frac{-q^{B,-}_N q^{B,+}  + q^{B,-} q^{B,+}_N }{\left|q_N^B\right|^2}\right]  = -\mathrm{Im\;}t_{\star}^B \;, \nn \\ 
\mathrm{Im\;}t_0^C &=& - \left[\frac{-q^{B,-}_N q^{B,+}  + q^{B,-} q^{B,+}_N 
- q^{A,-} q_N^{B,+}  + q^{A,+} q^{B,-}_N }{\left|q_N^B\right|^2} \right]  = -\mathrm{Im\;}t_{\star}^C 	\;.
\label{posicons}
\eea
We see that the modulus value on the wall where the central charge vanishes for the states $B$ and $C$  is exactly the negative of the attractor locus for those states. Since the wall is defined by (\ref{rtwssole2}) for any $\mathrm{Im\;}t > 0$, having no point on the wall of marginal stability where the central charge vanishes is the same as the statement that the two attractor loci are positive. This is precisely the problem we have analysed in section \ref{sec:nogo2elec}, and have shown that it leads to an obstruction.

Actually, the whole analysis is only valid for $\mathrm{Im\;}t \gg 1$, since that is required for the use of the nilpotent orbit approximation. 
%It is also required so that there is a path along the wall connecting the value of $\mathrm{Im\;}t$ at the point where it is crossed and the vanishing values (\ref{posicons}) where there are no singularities in the moduli space, and so we can apply the stability conjecture above. 
But the power of the stability constraint is that if we found an attractor locus which is very negative $\mathrm{Im\;}t_{\star} \ll -1$, then we can translate that into a zero of the central charge in the controlled region. The only remaining refuge for states could be in the strong-coupling regime $\left| \mathrm{Im\;}t\right| < 1$. So the only escape from the obstruction is if we pick charges such that both of $\mathrm{Im\;}t_0^B$ and $\mathrm{Im\;}t_0^C$ are either small or negative. However, looking at the constraint (\ref{nogo2c}) we see that it requires at least one of the attractor loci to be large and negative, which means at least one large and positive $\mathrm{Im\;}t_0$. 

The second powerful aspect of the stability constraint it that it applies on the wall of marginal stability. That means that the multi-parameter case can be analysed in the same way as the one-parameter case, as long as the degeneration itself is by variation of only one parameter. So the value of the $q$'s for the positivity constraints (\ref{posicons}), is the same as the one for the wall of marginal stability. There is no need to track them to the attractor loci. Their dependence on the other moduli $z^i$ is inconsequential for the constraints. 

There is an important way to attempt to avoid the problem of the vanishing central charge point $\mathrm{Im\;}t_0$. We can postulate that between the point on the wall of marginal stability which was crossed for the decay, and the vanishing central charge locus, the state decayed and left the spectrum. So it participated in the decay, and then after moving along the marginal stability locus, it decayed away so that it did not lead to a singularity in the moduli space at $\mathrm{Im\;}t_0$. It is difficult to believe that this could occur, because recall that the states have an associated BPS index which is typically very large, and this would have to completely vanish through decays upon the variation to $\mathrm{Im\;}t_0$. But we can be even more precise and show that in fact one can never avoid the singularity. Recall that we faced a similar possibility in section \ref{sec:nogoNelec} when studying attractor loci. We considered whether it is possible to avoid a negative attractor locus by having the state decay, so have a split attractor flow, to two states with positive attractor loci. We showed that this is not possible because one of the decay products would always have a negative attractor locus. We can import this result into this setting to show that if the state which has a vanishing point decays somewhere along the wall of marginal stability, then moving along the marginal stability wall of that decay will always lead to a zero locus for one of the products in a physical regime for $\mathrm{Im\;}t$. The only thing we need to show is that the correspondence between the attractor locus and the zero locus, as in (\ref{posicons}), is general. So the negative attractor locus which we cannot remove by decays turns into a positive vanishing point.\footnote{One might try to hide the vanishing of this second decay product in the strong-coupling region, but this is not possible. The simplest way to see this is that the analysis in section \ref{sec:nogoNelec} shows that having both $\left|\mathrm{Im\;}t_{\star}^B\right| \ll \left|\mathrm{Im\;}t_{WMS}\right|$ and $\left|\mathrm{Im\;}t_{\star}^C\right| \ll \left|\mathrm{Im\;}t_{WMS}\right|$ is not possible for $\mathrm{Im\;}t_{\star}^A < 0$. Therefore, at least one of $\mathrm{Im\;}t_{\star}^B$ or $\mathrm{Im\;}t_{\star}^C$ must be negative and large in magnitude.} 

It is simple to show that indeed the correspondence between the attractor loci and the vanishing loci is general. The general central charge, dropping the overall irrelevant Kahler-potential normalization, takes the form
\be
\mathrm{Re\;}Z\left(\bq\right) \sim q^+ + q_N^+ \; \mathrm{Re\;}t - q_N^- \;\mathrm{Im\;}t \;, \;\;
\mathrm{Im\;}Z\left(\bq\right) \sim q^- + q_N^+ \; \mathrm{Im\;}t + q_N^- \;\mathrm{Re\;}t \;. 
\label{rezimz}
\ee
For a charge participating in a decay, the wall of marginal stability can be defined by
\be
\mathrm{Re\;}Z\left(\bq\right) A\left(t\right) = B\left(t\right)\;\mathrm{Im\;}Z\left(\bq\right) \;,
\label{Adef}
\ee
where $A\left(t\right)$ and $B\left(t\right)$ are in general real functions of the moduli (in the case where one of the charges is purely electric it is a constant). Then if such a wall exists in a physical regime, we can combine (\ref{rezimz}) and (\ref{Adef}) to write that everywhere on the wall we have
\be
\mathrm{Re\;t}_{WMS} = \frac{-Bq^- + A q^+ - \left(A q_N^- + B q_N^+ \right) \mathrm{Im\;t} }{-Aq_N^+ + B q_N^-}\;.
\ee
Substituting this into (\ref{rezimz}) we can solve for the vanishing locus 
\be
\mathrm{Re\;}Z\left(\bq,t_0\right) = 0\;, 
\label{revanlo}
\ee
which gives
\be
\mathrm{Im\;}t_0 = -\left[\frac{-q^{-}_N q^{+}  + q^{-} q^{+}_N }{\left|q_N\right|^2}\right]  = -\mathrm{Im\;}t_{\star} \;.
\ee
The only subtlety here is if the vanishing locus is such that also $A\left(t_0\right)=B\left(t_0\right)=0$, but then we would have a vanishing locus (\ref{vansloc}) for one of the other states.

%%%%%%%%%%%%%%%%%%%%
\subsection{General proof of obstruction}
\label{sec:obsgen}
%%%%%%%%%%%%%%%%%%%%

It is possible to prove very generally, and relatively simply, the first part of the obstruction: that for electric to dyonic decays, one of the decay products must have a vanishing central charge point on the wall of marginal stability. In fact, one can prove a more general statement which we will outline below.

Let us begin by noting the reason why the solutions (\ref{posicons}) even exist. This is highly non-trivial because if we take a generic central charge, and look for its vanishing loci, they are two constraints on one coordinate and so pick out a point in moduli space. On the other hand, the wall of marginal stability is a line in the moduli space. And generic lines and points do not intersect. 

The point is that we are considering electric states for which the phase is a constant, so independent of $t$. The constraint for the wall of marginal stability (\ref{wmsZc1}) then ensures that the ratio of the real and imaginary parts of the central charges for $\bq^B$ (and $\bq^C$) are also a constant. They therefore become proportional to each other. This means that the requirement for the wall of marginal stability to have a vanishing point on it becomes the intersection of two real lines: the marginal stability line and the line of vanishing real part of the central charge. Once they intersect, the vanishing of the imaginary part of the central charge is guaranteed. In a real two-dimensional space, two generic real lines intersect.
%\footnote{Note that the wall of marginal stability ensures a relation between the real and imaginary parts of the central charge always. In the case when they are truly related by a constant they must share zeros. More generally, they naturally share zeros but not necessarily. It could be that on the locus where the real part of the central charge of state $B$ vanishes, the real part of the central charge for state $A$ vanishes instead of the imaginary part of the central charge for state $B$.} 

Since on the wall of marginal stability the real and imaginary parts of the central charge are proportional to each other, the second constraint for the decay (\ref{wmsZc2}) can be written as
\be
\frac{\mathrm{Re\;}Z\left(\bq^A\right)}{\mathrm{Re\;}Z\left(\bq^B,t_{WMS}\right)} \geq 1\;.
\label{rtzac}
\ee
It is important to note that in (\ref{rtzac}) we have manifested the point that the numerator is independent of $t_{WMS}$, it is a constant. The denominator is a function of $t_{WMS}$, but we only need (\ref{rtzac}) to hold for some point on the wall of marginal stability, so for some value of $t_{WMS}$.  

Now let us consider the vanishing point for the central charge of $\bq^C$, so $\mathrm{Im\;}t_0^C$. Since the marginal stability line spans all of $\mathrm{Im\;}t$, for this point not to lie one it we must have that there is no positive value of $\mathrm{Im\;}t$ for which $\mathrm{Re\;}Z\left(\bq^A\right)=\left.\mathrm{Re\;}Z\left(\bq^B,\mathrm{Im\;}t\right)\right|_{WMS}$. Since they must have the same sign to satisfy (\ref{rtzac}), we see that we must therefore have
\be
\left.\mathrm{Re\;}Z\left(\bq^B,\mathrm{Im\;}t\right)\right|_{WMS} > \mathrm{Re\;}Z\left(\bq^A\right) \mathrm{\;or\;\;} \left.\mathrm{Re\;}Z\left(\bq^B,\mathrm{Im\;}t\right)\right|_{WMS} < \mathrm{Re\;}Z\left(\bq^A\right)\;\;\; \forall \;\mathrm{Im\;}t \;.
\label{2optio}
\ee
But $\left.\mathrm{Re\;}Z\left(\bq^B,\mathrm{Im\;}t\right)\right|_{WMS}$ is unbounded in magnitude as a function of $\mathrm{Im\;}t$, and therefore only the first option in (\ref{2optio}) is possible. Therefore, (\ref{rtzac}) can never be satisfied. 

%In fact, we can see now directly from (\ref{rtzac}) why such a condition cannot be satisfied. The point is that while we may imagine crossing the wall of marginal stability at a particular point, so a particular value of $\mathrm{Im\;}t$, whether the decay occurs or not should not depend on the choice of this point. This is essentially what the analysis around the stability conjecture above are capturing. Since the wall of marginal stability is a line such that the denominator of (\ref{rtzac}) is unbounded as a function, the constraint (\ref{rtzac}) can never hold everywhere on the wall. 

The analysis presented generalises simply since it does not rely on a particular choice of $n$ or $d$. The only assumption was that electric charges are such that $\mathrm{Re\;}Z\left(\bq^A\right)$ is a constant (up to the overall factor from the Kahler potential). 
%Indeed, it also makes clear that it is part of a more general statement, which we discuss in the next section. 
We can also see that this restriction is simple to generalise. The absence of a zero locus along the wall of marginal stability is the same as the statement that (\ref{rtzac}) must hold everywhere on the wall of marginal stability. 

It is worth showing this explicitly. Let us consider in general a decay $A \rightarrow B + C$. Then we require the generalisation of (\ref{rtzac}), so
\be
\left. \frac{\mathrm{Re\;}Z\left(\bq^A,t_{c}\right)}{\mathrm{Re\;}Z\left(\bq^B,t_{c}\right)} \right|_{WMS} \geq 1\;.
\label{rtzacgen}
\ee
Here we made things more explicit than (\ref{rtzac}) by labelling explicitly the point on the wall of marginal stability where the decay is occurring as $t_c$. Now let us violate (\ref{rtzacgen}) at some point by moving along the wall of marginal stability, so that eventually we reach some point $t_0$ such that 
\be
\left. \frac{\mathrm{Re\;}Z\left(\bq^A,t_{0}\right)}{\mathrm{Re\;}Z\left(\bq^B,t_{0}\right)} \right|_{WMS} = 1\;.
\label{rzrvio}
\ee
At this point we have
\be
\mathrm{Re\;}Z\left(\bq^C,t_{0}\right) = \mathrm{Re\;}Z\left(\bq^A,t_{0}\right) - \mathrm{Re\;}Z\left(\bq^B,t_{0}\right) = 0 \;.
\ee
But we also know that everywhere on the wall of marginal stability we have
\bea
\mathrm{Re\;}Z\left(\bq^C,t\right) \mathrm{Im\;}Z\left(\bq^A,t\right) &=& \mathrm{Re\;}Z\left(\bq^A,t\right) \mathrm{Im\;}Z\left(\bq^C,t\right) \;, \nn \\
\mathrm{Re\;}Z\left(\bq^C,t\right) \mathrm{Im\;}Z\left(\bq^B,t\right) &=& \mathrm{Re\;}Z\left(\bq^B,t\right) \mathrm{Im\;}Z\left(\bq^C,t\right) \;.
\eea
So at $t_0$ we have only two possibilities
\be
\mathrm{Re\;}Z\left(\bq^C,t_0\right) = \mathrm{Im\;}Z\left(\bq^C,t_0\right) = 0 \;, 
\label{op1}
\ee
or
\be
\mathrm{Re\;}Z\left(\bq^A,t_0\right)  = \mathrm{Re\;}Z\left(\bq^B,t_0\right)  = \mathrm{Re\;}Z\left(\bq^C,t_0\right)  = 0 \;.
\label{op2}
\ee
The second option (\ref{op2}) is not viable because $t_0$ is defined by (\ref{rzrvio}). It is possible to consider the possibility that in (\ref{rzrvio}) the numerator and denominator share a zero at $t_0$ while leaving a finite ratio, but this seems highly implausible since they are different functions in one real variable. In other words, the wall of marginal stability guarantees that $\mathrm{Re\;}Z\left(\bq^C,t_0\right)$ is proportional to $\mathrm{Im\;}Z\left(\bq^C,t_0\right)$, but not that $\mathrm{Re\;}Z\left(\bq^A,t_0\right)$ is proportional to $\mathrm{Re\;}Z\left(\bq^B,t_0\right)$. We are therefore left with the first option (\ref{op1}), and so $Z\left(\bq^C\right)$ vanishes somewhere on the wall of marginal stability. 

Yet another, the most simple, way to see this is to note that (\ref{rzrvio}) is the statement that state $A$ and $B$ have equal masses. But the wall of marginal stability is precisely the locus where the masses of the constituents, $B$ and $C$, sum up to the mass of $A$. So state $C$ must be massless there.

There is an important subtlety with connecting this to the $g \rightarrow 0$ limit, specifically, it is not always the case that the wall of marginal stability stretches to the degeneration limit $\mathrm{Im\;t} \rightarrow \infty$. In the $n=d=1$ case, for an electric to dyonic decay, it did because the condition defining the wall of marginal stability (\ref{wmsZc1}) was a linear equation in $\mathrm{Im\;t}$, and so has solutions for all its values. On the other hand, the wall for a dyonic state to decay to a dyonic state led to a quadratic equation which for fixed charges had no solution as $\mathrm{Im\;t} \rightarrow \infty$. 

In order to have a solution for arbitrarily large $\mathrm{Im\;t}$ we need to look for a solution scaling of type 
\be
\mathrm{Re\;t} = x \;\mathrm{Im\;t} \;,\;\; x \in \mathbb{R} \;.
\label{cndx}
\ee
Whether such a solution exists for a real $x$ depends on the highest power of $t$ which appears in the central charge for the states $A$ and $B$, and is shown in table \ref{tab:sol}. We see that if the power of $t$ is larger in state $B$ than $A$, the wall of marginal stability stretches out to  $\mathrm{Im\;t} \rightarrow \infty$. So let us assume this for now, and we will explain how this highest power is classified mathematically in section \ref{sec:decayfilt}. With this assumption, taking the degeneration limit $\mathrm{Im\;t} \rightarrow \infty$, which corresponds to $g \rightarrow 0$, gives the BPS Stability Filtration of the introduction. 
\begin{table}
\center
\begin{tabular}{|c||c|c|c|c|}
\hline
{\bf A / B} & {\bf 0} & {\bf 1} & {\bf 2} & {\bf 3} \\ 
\hline
\hline
{\bf 0} & - & \checkmark & \checkmark & \checkmark \\
\hline
{\bf 1} & \checkmark & $\times$ & \checkmark & \checkmark\\
\hline
{\bf 2} & \checkmark & \checkmark  & $\times$ & \checkmark \\
\hline
{\bf 3} & \checkmark  & \checkmark & \checkmark  & $\times$ \\
\hline
\end{tabular}
\caption{Table showing if a solution of type (\ref{cndx}) exists, and therefore the wall of marginal stability stretches to the limit $\mathrm{Im\;t} \rightarrow \infty$ (and so $g \rightarrow 0$), for the highest powers of $t$ in the (holomorphic part of the) central charge of states $A$ and $B$. When a solution exists, there are no constraints on the charges.}
\label{tab:sol}	
\end{table}

The reason that the BPS Stability Filtration is a proposal and not proven is that there remains the issue of showing that it is not possible for the state with the vanishing locus to participate in the decay, and then decay away from the spectrum by the time it reaches the vanishing point. For the case $n=d=1$ we have proven that indeed this cannot happen, or more precisely, that it will lead to some other decay product having a zero locus. We have not done so for general $n$ and $d$, which is the missing step for a general proof. Unfortunately, showing this requires some further work, which we leave for the future. We expect that this is only technical and not conceptual, the accumulated evidence so far seems sufficiently strong to propose a BPS Stability Filtration. 

We can make some simple statement about the general case by restricting to certain charge sectors. In general, the central charge takes the form
\bea
Z\left(\bq,t,z^i\right) &=& \frac{1}{\left| c\left(z^i\right)\mathrm{Im\;}t \right|^{\frac{d}{2}}} \left[ \left<\bq,\ba_0\left(z^i\right)\right>  + \left<\bq,N\cdot\ba_0\left(z^i\right)\right> t + \right. \nn \\
& & \frac12\left.\left<\bq,N^2\cdot\ba_0\left(z^i\right)\right> t^2 + \frac16\left<\bq,N^3\cdot\ba_0\left(z^i\right)\right> t^3 \right] \;. 
\label{Ztd3}
\eea
We can choose our black hole charges such that
\be
\left<\bq,\ba_0\left(z^i\right)\right> = \left<\bq,N^3\cdot\ba_0\left(z^i\right)\right> = 0 \;.
\ee
We then define
\be
\tilde{\bq} = N \cdot \bq \;.
\ee
so that the central charge takes the form (for $d=3$)
\be
Z\left(\bq,t,z^i\right) = \frac{t}{\left| c\left(z^i\right) \mathrm{Im\;}t\right|^{\frac{3}{2}}} \left[\left<\tilde{\bq},\ba_0\left(z^i\right)\right> + 
\frac12\left<\tilde{\bq},N\cdot\ba_0\left(z^i\right)\right> t \right]  \;. 
\label{Ztd3red}
\ee
The analysis of the obstruction to decays from the electric to dyonic charges is insensitive to overall factors in the central charge. Therefore, this is essentially the same form as (\ref{Ztd1n1}) and we can carry our results over.

It is worth discussing the relation of this analysis to that of \cite{Andriyash:2010yf}. There, paths in moduli space which connect marginal stability walls to anti-marginal stability walls (where the central charges anti-align) were studied. In our setting, the locus where the central charge of one of the constituents vanishes turns the marginal stability wall into an anti-marginal stability wall. So one can consider circling that locus, thereby traversing a path as studied in \cite{Andriyash:2010yf}. There it was argued that either there is a singularity in the moduli space (which we are forbidding) that induces a conjugation wall, or the constituents themselves were never fundamental in the first place, and their constituents rearrange along the path (this was termed crossing a recombination wall). This is consistent with our analysis. Of course, the assumption of \cite{Andriyash:2010yf} is that BPS constituents are populated in the first place, which need not be the case here. 

%%%%%%%%%%%%%%%%%%%%
\subsection{A BPS Stability Filtration}
\label{sec:decayfilt}
%%%%%%%%%%%%%%%%%%%%

The analysis in section \ref{sec:obsgen} culminated in the point that if we consider a decay of some state $A$ to some state $B$ (plus something else), then the decay requires that the mass of $A$ is larger than the mass of $B$ over the full locus of marginal stability. We also showed that if the highest power of $\mathrm{Im\;}t$ in the central charge of $B$ is higher than the highest power of $\mathrm{Im\;}t$ in the central charge of $A$, then the wall of marginal stability stretches to the limit $\mathrm{Im\;t} \rightarrow \infty$ (and so $g \rightarrow 0$). This also automatically implies that for some sufficiently large value of $\mathrm{Im\;t}$ the mass of $B$ will become higher than the mass of $A$, and so is a sufficient condition to forbid such a decay.

There is a mathematical formulation which captures precisely the highest power of $\mathrm{Im\;}t$ in the central charge, termed the monodromy weight filtration. Here we outline some parts of it, and refer to \cite{Grimm:2018ohb,Grimm:2018cpv} for much more detail about this structure.

Consider the rational vector space of charges $\bq \in V\left(\mathbb{Q}\right)$. Then the monodromy matrix $N$ induces a monodromy weight filtration $W_i\left(N\right)$ on $V$.
\be
W_{-1}\equiv 0\ \subset\  W_0\ \subset\ W_1\ \subset\ ...\  W_{6} = V\ .
\label{filtration}
\eeq
This filtration is uniquely specified through the following defining properties
\begin{align}
 &  \bullet\quad N W_i \subset W_{i-2} & \label{Nw2} \\
   &\bullet \quad  N^j : Gr_{3+j} \rightarrow Gr_{3-j}\ \ \text{is an isomorphism,}  \;\;\;\;Gr_{j} \equiv W_{j}/W_{j-1}\ .   
\end{align}
Note that there is a simple representation of the $W_i$ in terms of the kernels and images of $N^j$ as 
\be
   W_0 = \text{im}\,N^3\ ,\quad W_{1} = \text{im}\, N^{2} \cap \text{ker}\,N\ ,\ \ldots \ ,\quad
    W_{5} = \text{ker} N^3\ .
\ee
This implies immediately that if the unipotency index is smaller than the complex dimension of the manifold, $n<3$, some of the previous subspaces will be empty. In particular,  for all $j>n$  we have $W_{3+j}=W_{3+n}$ and $W_{3-j}=0$. 

Note that using the uniqueness of the filtration on can show that if we have $\bq_A \in W_i$ and $\bq_B \in W_{6-i-j}$ then 
\beq 
\label{orthog}
   \left<\bq_A,\bq_B\right> =0\;, \mathrm{if\;} j>0 \ .
\eeq

The splitting into electric and magnetic states can be done as follows. We define electric states as those lying in $W_i$ with $i < 3$. Dyonic states are charges lying in $W_i$ with $i>3$. Magnetic states are dyonic states where the electric components, so those that are in $W_i$ with $i<3$, are removed. The simplest way to identify those is by acting with $N$ and using (\ref{Nw2}).  

The $W_i$ are related to the highest power of $t$ which appears in the (holomorphic part of the) central charge. This is because (from (\ref{Nw2}))
\be
\left<\bq,N^w\cdot \ba_0 \right> = \left(-1\right)^w\left<N^w\cdot\bq,\ba_0 \right> \neq 0 \implies \bq \notin W_{2\left( w+1 \right) - n} \;.
\ee
When using this we should keep in mind that $W_{4+n}$ does not exist, and $W_{2-n}$ is empty. We can therefore assign an {\it s-weight} to a charge as
\be
\mathrm{s-weight}\left[\bq\right] = \mathrm{Largest\;integer\;}w\;\mathrm{such\;that\;} \bq \notin W_{2\left( w+1 \right) - n} \;.
\label{weigdef}
\ee
Then the s-weight of a BPS state corresponds to the highest allowed power of $t$ in its central charge.

We therefore can identify the BPS stability filtration with the monodromy weight filtration as follows: a BPS state cannot decay to a BPS state of a higher s-weight. The remarkable thing is that this is true for any perturbative $g \ll 1$, and not only in the $g \rightarrow 0$ limit.

There is a subtlety with such an identification of the BPS stability filtration with the monodromy weight filtration which is that it assumed the the maximum allowed power of $t$ is realised by the charges. This assumes that the charge is a sufficiently generic element of the (maximal) $W_i$ that it belongs to. A better way to phrase this genericity condition is that the moduli other than $t$, so the $z^i$, are at generic values. In that case a vanishing $\left<\bq,N^w\cdot \ba_0 \right>$ implies the full constraint on the charges in terms of the monodromy weight filtration.

If we would like to drop the assumption of generic values for the other moduli, then there is still a mathematical formulation which relates the highest power of $t$ which appears in the central charge to the charge $\bq$, but it is slightly more involved. It corresponds to placing the charge $\bq$ into the primitive spaces of the limiting mixed Hodge structure (the $P_i$ in \cite{Grimm:2018ohb,Grimm:2018cpv}). On those spaces there exists a polarized inner product, which would guarantee the non-vanishing of inner products of type $\left<\bq,N^w\cdot \ba_0 \right>$. Indeed, the most precise way to formulate the stability constraints is by using the Deligne splittings in a basis where the $Sl_2$-algebras associated to the different $N_i$ are fully commuting. Such a Deligne splitting is so-called $\mathbb{R}$-split and so any real charge can be placed into the space of a Deligne sub-space plus its conjugate. The process of reaching such a basis, and the properties of the splitting, is explained in detail in \cite{Grimm:2018cpv}.\footnote{Similarly, the identification with the monodromy weight filtration assumes $n=d$, which to our knowledge is correct for any infinite distance locus, but can again be removed as an assumption by going to an $\mathbb{R}$-split Deligne basis.}

%%%%%%%%%%%%%%%%%%%%
\section{Summary and discussion}
\label{sec:sum}
%%%%%%%%%%%%%%%%%%%%

In this paper we studied the decay and spectrum of BPS states in ${\cal N}=2$ theories, in particular considering weak-coupling limits in moduli space $g \rightarrow 0$. We found evidence for a certain filtration structure in BPS states, which is determined by their mass in the $g \rightarrow 0$ limit. Specifically, that a BPS state at any (perturbative) value of $g$ cannot decay to a BPS state whose mass at $g \rightarrow 0$ is infinitely larger. This appears somewhat surprising, for example, an electrically charged black hole cannot emit a dyonic particle upon crossing a wall of marginal stability in any perturbative regime, because in the $g \rightarrow 0$ limit the black hole would be infinitely lighter than the dyon.

Although the analysis was sometimes involved, at least some of the key points are extremely simple and so we can summarise them here using quite general notation. Consider a decay of states $A \rightarrow B + C$. Let this decay occur at vanishing kinetic energy, like in ${\cal N}=2$. Let the masses of the states depend on some couplings, say $g$ for simplicity. Then at the decay point in coupling space, say $g=g_c$, we have an equality in the masses $M_A\left(g_c\right) = M_B\left(g_c\right) + M_C\left(g_c\right)$. Now consider this equality of the masses as a constraint on the value of the couplings. It picks out a real co-dimension one subspace of the coupling space. In ${\cal N}=2$ this is a wall of marginal stability. Now let this subspace be non-compact, so that it stretches out to the limit $g \rightarrow 0$. This is sometimes, not always, true in ${\cal N}=2$. Then we see that if somewhere between $g=g_c$ and $g \rightarrow 0$ the ordering of the masses changes, so say $M_B$ becomes larger than $M_A$, we must have that one of the constituents of the decay becomes massless, in this case $M_C=0$. A massless charged state induces a singularity in the low-energy effective theory, so in ${\cal N}=2$ would manifest as a singularity in the moduli space. If we insist on the absence of such a massless state, or singularity, then we forbid a change in the ordering of the masses.
This induces a certain filtration on the decay of states which is independent of $g$, so in particular can be deduced at $g \rightarrow 0$. 

The absence of a singularity in perturbative regimes in ${\cal N}=2$ is guaranteed by the nilpotent orbit theorem, which gives the universal behaviour around any infinite distance (or weakly-coupled) locus in moduli space.\footnote{This gives also a rather precise statement of what we mean by the perturbative regime: it is one where instantons are suppressed. Note that we do not consider cases where $g \rightarrow 0$ at finite distances in moduli space, such as the conifold locus, as perturbative regimes. The vanishing coupling there is only due to integrating out a massless state, and would be finite at any finite mass scale.} The only remaining subtlety is that it could be that state $C$ participated in the decay process, and then as we varied the coupling, itself decayed away so that it was not present in the spectrum at $M_C=0$. This seems a contrived situation to occur, but we need to rule it out to be sure. We can prove that this is not possible in certain (one of the three possible) weak coupling limits. More precisely, we showed that if it does decay, then at least one of the constituents would become massless in the perturbative regime still leading to a singularity in the moduli space. So the singularity is robust against decays in this sense. We were not able to prove this for all possible weak-coupling limits, simply because the algebraic equations are of order $2d$, with $d=1,2,3$ being the different possible limits (we solved $d=1$, so the quadratic case). But we believe it is a technical task to show it generally. This belief led to the proposal of a BPS Stability Filtration, as stated in the introduction.

We gave further evidence, and developed the understanding around the filtration, by studying also attractor flows for multi-centred black holes. These should capture the spectrum of BPS states, and we find agreement with the filtration proposal. So a decay which would violate the filtration, manifested as a split attractor flow, leads to an unphysical attractor locus for one of the constituents. Also, the filtration can be understood as the statement that the wall crossing formula \cite{Kontsevich:2008fj} should apply to all states in the theory, everywhere on the wall (so imposing the absence of non-stable states anywhere on the wall). 

The filtration on the BPS states can be mathematically formulated in terms of the Monodromy Weight Filtration associated to the limiting mixed Hodge structure of the $g \rightarrow 0$ limit. The statement is that a state cannot decay to one which has a higher s-weight (where s-weight is defined in (\ref{weigdef})). This condition also automatically captures the requirement that the wall of marginal stability stretches to the $g \rightarrow 0$ point: it is true whenever the s-weights of the decaying state and one of its products differ. Of course, since BPS states are mapped to the existence of certain geometric sub-manifolds such as special-Lagrangians, the filtration implies a corresponding geometric statement which is mathematically interesting.

It is natural to propose that our analysis can be extended to all $g$, so away from the perturbative regime, by assuming the absence of moduli space singularities on the walls of marginal stability. So in this sense, it is the statement that the stability of BPS states depends only on the singularities in the moduli space, whether they are at finite or infinite distance. 

If, as we propose, a filtration exists on BPS states in ${\cal N}=2$ theories, then what does this imply for the Swampland motivations discussed in the introduction? Regarding the first topic, the black hole to particle transition as $g \rightarrow 0$. Electrically charged black holes, if we can make sense of their divergent attractor flows (see below), cannot emit any dyons, and therefore all their constituents are completely stable throughout the transition. The tower of light states as $g \rightarrow 0$ therefore start their life as black hole microstates. Note that this is somewhat expected from the fact that the number of BPS states in examples grows exponentially already from small charges. In spirit, this goes against the idea of the electric weak gravity conjecture which would like the black holes to discharge so as to avoid stable charged remnants. Of course, as discussed in the introduction, the ${\cal N}=2$ BPS decays are very different to the discharge decays of the weak gravity conjecture, so we must be quite careful in extrapolating any connections. 

 Regarding the second topic of constructing multi-centre flows to understand electric black holes which exhibit divergent single-centre attractor flow. A filtration implies that no such flows can exist in any perturbative region of moduli space (something which we checked in multiple examples). This has a natural interesting interpretation once we match the split attractor flows onto some kind of resolution of the constituents of states. It means that the filtration captures some sort of 'fundamentality' of the states, so states may decay to more fundamental ones, but not the other way around.

The final motivation in the introduction was to understand the emergence proposal, and the related idea of moduli space holography \cite{Grimm:2020cda,Grimm:2021ikg}. What we find is essentially evidence for (at least some version of) these ideas. The decay, and therefore the spectrum, of BPS states in the `bulk', at finite $g$, is determined by their behaviour at the `boundary' $g \rightarrow 0$. This is a notion of holography. We find also a relation between attractor flows and the vanishing mass obstructions, and attractor flows are a manifestation of some sense of moduli space non-locality (since the spectrum at the moduli values at spatial infinity is determined by the the moduli attractor values). In this sense, the analogies with holography are interesting. 

We note also that the idea that the filtration is associated with how `fundamental' the states are matches nicely with the emergence proposal, where the tower of light charged states are thought of as constituents that the gauge field is composed of. 

More generally, we find the results quite surprising in that they make the $g \rightarrow 0$ limit much more powerful than one might naively guess. This is promising in terms of the Swampland program, which aims to constrain and understand such limits.

\noindent {\bf Acknowledgements:} This research was supported by the Israel Science Foundation (grant No. 741/20).

\appendix

%%%%%%%%%%%%%%%%%%%%%%%%%%%%%%%%%%%%%%%%%%%%%%%
\section{Example one-parameter Calabi-Yau}
\label{sec:oneparamexam}
%%%%%%%%%%%%%%%%%%%%%%%%%%%%%%%%%%%%%%%%%%%%%%%%

Throughout the paper we have studied one-parameter $d=1$ limits in much detail. While our approach has usually been completely general, it is worth having in mind an explicit example. In this appendix we present such an example as a guide to the more general results in the main text. Many of the details of the analysis follow that of \cite{Joshi:2019nzi}. 

\begin{figure}
\centering
 \includegraphics[width=1.0\textwidth]{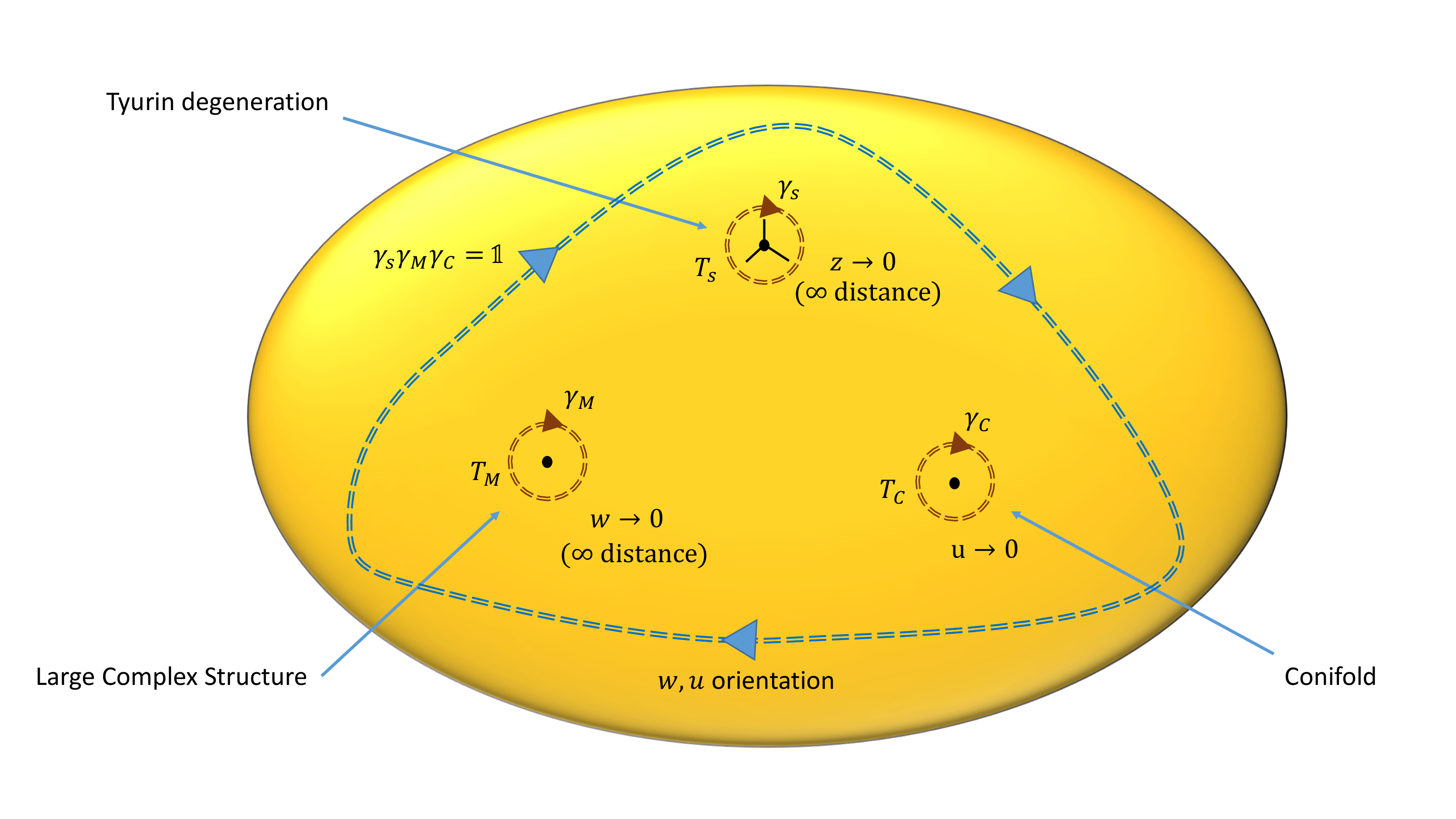}
\caption{Figure showing the moduli space of the $\mathbb{P}^5[3,3]$ Calabi-Yau. The orientation of the different monodromies, $\gamma_s$, $\gamma_M$ and $\gamma_C$, about the singular loci is shown. The global orientation ensures that the product of the monodromies is the unit operator.}
\label{fig:modulispace}
\end{figure}

The example we take is the bi-cubic Calabi-Yau $\mathbb{P}^5[3,3]$. It is specified as 
\bea
& & \frac{x_1^3}{3} + \frac{x_2^3}{3} + \frac{x_3^3}{3} + \psi x_1 x_2 x_3 = 0 \;, \nn \\
& & \frac{x_4^3}{3} + \frac{x_5^3}{3} + \frac{x_6^3}{3} + \psi x_4 x_5 x_6 = 0 \;.
\eea
Where the $x_i$ are homogeneous coordinates in $\mathbb{P}^5$. The complex structure moduli space is spanned by the single complex parameter $\psi$. In terms of the vector multiplets we have $n_v=1$. It is useful to introduce coordinates on the moduli space 
\be
z \equiv \psi^2 \;,\;\; w \equiv \frac{1}{\mu \psi^6} \;,\;\; u \equiv 1-w \;,
\label{codef}
\ee
where $\mu = 3^6$. The moduli space has three special points
\begin{itemize}
\item Large complex structure : $w = 0$,
\item Tyurin degeneration : $z = 0$,
\item Conifold point: $u = 0$.
\end{itemize}
The first two of these are at infinite distance, while the conifold is at finite distance. The moduli space is therefore of the form $\mathbb{P}^1/\left\{ 0,1,\infty\right\}$ and is illustrated in figure \ref{fig:modulispace}. 

The Tyurin degeneration locus $z=0$ has a $\mathbb{Z}_3$ orbifold symmetry which can be seen by noting that sending $\psi \rightarrow \psi e^{\frac{2 \pi i}{3}}$ can be undone by rotatining the coordinates appropriately. This means that while the natural holomorphic coordinate for the conifold and large complex-structure behaves as $\psi^6$, the natural coordinate for the Tyurin degeneration is $\psi^2$. This explains the definition (\ref{codef}). 

There is a global orientation of the monodromies on the moduli space. As shown in figure \ref{fig:modulispace}, we take this such that we are circling all the monodromy loci clockwise in the coordinates $w$ and $u$. This means that in terms of the coordinates $z$ we are circling the Tyurin degeneration locus anti-clockwise. So if we label the monodromy loci by $M$ for the large complex structure point, $C$ for the conifold, and $s$ for the Tyurin degeneration, then the monodromy matrices associated to each locus are defined such that
\be
\bP_M\left(w e^{-2\pi i} \right) = T_M \cdot \bP_M\left(w\right) \;, \;\; \bP_C\left(u e^{-2\pi i} \right) = T_C \cdot \bP_C\left(u\right) \;,\;\; \bP_s\left(z e^{2\pi i} \right) = T_s \cdot \bP_s\left(z\right) \;.
\ee
Where we denoted the local form of the period vector around the monodromy locus with the appropriate subscript.

In this section we will first give expressions for the period vector and monodromy matrices associated to the three points of interest. It is important to note that the monodromy matrices are not in a common symplectic basis. This means, for example, that their product does not give the unit matrix. This part of the section builds heavily on the analysis in \cite{Joshi:2019nzi}. In the second part we will use the period vector to study the spectrum of BPS states near the monodromy loci. 

%%%%%%%%%%%%%%%%%%%%
\subsection{The period vector}
%%%%%%%%%%%%%%%%%%%%

In this subsection we calculate the local expressions for the period vector about the monodromy loci. We are interested in the $n=d=1$ locus (Tyurin degeneration). We give the expressions for the conifold locus for completeness in appendix \ref{sec:conlcslo}. 

There are two bases of interest for the period vector, the Frobenius basis $\tilde{\bP}$ and the symplectic basis $\bP$. The Frobenius period vector can be expressed in terms of Meijer G-functions as \cite{Joshi:2019nzi}
\be
\tilde{\bP}_s=\left( \begin{array}{c} 
G_{4,4}^{1,4}\left( -\psi^6 \; | \begin{array}{cccc} 1 & 1 & 1 & 1 \\ \frac13 & \frac13 & \frac23 & \frac23 \end{array} \right) \\
G_{4,4}^{1,4}\left( -\psi^6 \; | \begin{array}{cccc} 1 & 1 & 1 & 1 \\ \frac23 & \frac13 & \frac13 & \frac23 \end{array} \right) \\
G_{4,4}^{2,4}\left( \psi^6 \; | \begin{array}{cccc} 1 & 1 & 1 & 1 \\ \frac13 & \frac13 & \frac23 & \frac23 \end{array} \right) \\
G_{4,4}^{2,4}\left( \psi^6 \; | \begin{array}{cccc} 1 & 1 & 1 & 1 \\ \frac23 & \frac23 & \frac13 & \frac13 \end{array} \right) \\
\end{array} \right) \;.
\ee
We can expand these about $\psi=0$ to give
\be
\tilde{\bP}_s=\left( \begin{array}{c} 
\psi^2 \left( -1 \right)^{\frac13} \left( \frac{\Gamma\left(\frac13 \right)}{\Gamma\left(\frac23 \right)}\right)^2  + \psi^8 \left( -1 \right)^{\frac13} \left( \frac{\Gamma\left(\frac13 \right)^2}{6\Gamma\left(\frac23 \right)}\right)^2   + ... \\
\psi^4 \left( -1 \right)^{\frac23} \left( \frac{\Gamma\left(\frac23 \right)^2}{\Gamma\left(\frac43 \right)}\right)^2  + \psi^{10} \left( -1 \right)^{\frac23} \left( \frac{\Gamma\left(\frac23 \right)^2}{3\Gamma\left(\frac43 \right)}\right)^2   + ... \\
-\psi^2 \left( \frac{\Gamma\left(\frac13 \right)^2}{\Gamma\left(\frac23 \right)}\right)^2 \Big(2\gamma + 6 \log \psi + 4 \Gamma\left(0,\frac13 \right) - 2\Gamma\left(0,\frac23 \right) \Big) + ... \\
-\psi^4 \left( \frac{\Gamma\left(\frac23 \right)^2}{\Gamma\left(\frac43 \right)}\right)^2 \Big(2\gamma + 6 \log \psi + 4 \Gamma\left(0,\frac23 \right) - 2\Gamma\left(0,\frac43 \right) \Big) + ... 
\end{array} \right) \;.
\label{Pitex}
\ee
Note that the overall prefactor of $\psi^2$ can be removed by a Kahler transformation and will not feature in the physically relevant quantities. 

The monodromy matrix in the Frobenius basis is defined as 
\be
\tilde{\bP}_s\left( z e^{2\pi i} \right)= \left( \mathbb{1} + \tilde{N}_s \right) \tilde{\bP}_s\left( z \right) \;.
\ee
From (\ref{Pitex}) we then obtain
\be
\tilde{N}_s = -6 \pi i\left( \begin{array}{cccc} 
0 & 0 & 0 & 0 \\
0 & 0 & 0 & 0 \\
\left( -1 \right)^{-\frac13}  & 0 & 0 & 0 \\
0 & \left( -1 \right)^{-\frac23}  & 0 & 0 
\end{array}\right) \;.
\ee

We can move to a more convenient symplectic basis which matches the general analysis in this work more directly through a transformation matrix
$U_{Ms}$ as
\be
\bP_s= U_{Ms} \cdot \tilde{\bP}_s \;.
\ee
The form of $U_{Ms}$ is \cite{Joshi:2019nzi}
\begin{equation}
	U_{Ms}=\left(
	\renewcommand{\arraystretch}{1.4}
	\begin{array}{cccc}
	 -\frac{9 \left(3+i \sqrt{3}\right)}{8 \pi ^2} & \frac{9 \left(3-i \sqrt{3}\right)}{8 \pi ^2} & \frac{9 i}{8 \pi ^3} & \frac{9 i}{8 \pi ^3} \\
	 -\frac{3+i \sqrt{3}}{8 \pi ^2} & \frac{3-i \sqrt{3}}{8 \pi ^2} & \frac{21 \left(1-i \sqrt{3}\right)}{8 \left(\sqrt{3}-5 i\right) \pi ^3} & \frac{3 i \left(4
	   \sqrt{3}+i\right)}{4 \left(\sqrt{3}-5 i\right) \pi ^3} \\
	 -\frac{3 \left(5+i \sqrt{3}\right)}{8 \pi ^2} & \frac{3 \left(5-i \sqrt{3}\right)}{8 \pi ^2} & \frac{3 \left(\sqrt{3}+3 i\right)}{16 \pi ^3} & -\frac{3 \left(\sqrt{3}-3
	   i\right)}{16 \pi ^3} \\
	 -\frac{3+i \sqrt{3}}{4 \pi ^2} & \frac{3-i \sqrt{3}}{4 \pi ^2} & \frac{3 i}{8 \pi ^3} & \frac{3 i}{8 \pi ^3} \\
	\end{array}
	\right)\;.
	\label{Tmat}
\end{equation}
We can translate the monodromy matrix into the symplectic basis by using (\ref{Tmat}) through
\be
N_s' = U_{Ms} \cdot \tilde{N}_s \cdot U_{Ms}^{-1} \;,
\ee
which gives
\be
N_s' = \left( \begin{array}{cccc} 
9 & 6 & -18 & -3 \\
6 & 1 & -3 & -14 \\
6 & 3 & -9 & -6 \\
3 & 2  & -6 & -1 
\end{array}\right) \;.
\ee
It is useful to perform a further symplectic rotation 
\be
N = M \cdot N_s \cdot M^{-1} \;,
\ee
with 
\be
M = \left( \begin{array}{cccc} 
4 & 3 & -9 & 0 \\
1 & 0 & 0 & -3 \\
\frac{5}{58} & -\frac{1}{87} & \frac{1}{29} & \frac{5}{174} \\
-\frac{1}{29} & \frac{53}{174} & \frac{5}{58} & -\frac{1}{87} 
\end{array}\right) \;.
\label{Msym}
\ee
This yields
\be
N = \frac19 \left( \begin{array}{cccc} 
0 & 0 & 0 & 0 \\
0 & 0 & 0 & 0 \\
2 & 1 & 0 & 0 \\
1 & 14 & 0 & 0 
\end{array}\right) \;.
\ee
Note that such a symplectic transformation to a form where $N$ only has components in a $2\times2$ block is always possible \cite{GGK}.

We can write the period vector in the canonical Nilpotent orbit form as
\be
\tilde{\bP}_s\left( z \right)  = e^{\frac{\tilde{N}_s \log z}{2\pi i}} \tilde{\ba}_s\left( z \right) \;,
\ee
where $z$ is defined as in (\ref{codef}). 
The holomorphic vector takes the form
\be
\tilde{\ba}_s\left(z \right) = z \left( \tilde{\ba}_{s,0} + z \;\tilde{\ba}_{s,1} \right) + {\cal O}\left( z^{5}\right) \;,
\label{batyuz}
\ee
with the expressions for the vectors
\be
\tilde{\ba}_{s,0} = \left( \begin{array}{c}  \frac{\left(-1\right)^{\frac13} \Gamma\left(\frac13 \right)^4}{\Gamma\left(\frac23 \right)^2} \\
0 \\
-2 \frac{\Gamma\left(\frac13 \right)^4}{\Gamma\left(\frac23 \right)^2} \left(\gamma + 2 \Gamma\left(0,\frac13 \right) - \Gamma\left(0,\frac23 \right) \right) \\
0
\end{array}\right) \;,\;\; \tilde{\ba}_{s,1} = \left( \begin{array}{c} 0 \\  \frac{\left(-1\right)^{\frac23} \Gamma\left(\frac23 \right)^4}{\Gamma\left(\frac43 \right)^2} \\
0 \\
-2 \frac{\Gamma\left(\frac23 \right)^4}{\Gamma\left(\frac43 \right)^2} \left(\gamma + 2 \Gamma\left(0,\frac23 \right) - \Gamma\left(0,\frac43 \right) \right) 
\end{array}\right) 
\ee
We can also define the symplectic limiting vector
\be
\ba_s'\left( z \right) \equiv U_{Ms} \cdot \tilde{\ba}_s\left( z \right)  \;,
\ee
and write
\be
\bP_s'\left( z \right) = e^{\frac{N'_s \log z}{2\pi i}} \ba'_s \left( z \right) \;.
\ee
Finally we should define the appropriately rotated vector
\be
\ba\left(z \right) = M \cdot \ba'_s\left(z \right) \;.
\ee
We have that $\ba_{0}$ is given explicitly by
\be
\ba_{0} = 
\left( \begin{array}{c}  
\frac{3\left(9+i\sqrt{3}\right) \Gamma\left(\frac13 \right)^4}{8\pi^2\Gamma\left(\frac23 \right)^2} \\
-\frac{3i\sqrt{3} \Gamma\left(\frac13 \right)^4}{4\pi^2\Gamma\left(\frac23 \right)^2}  \\
\frac{\Gamma\left(\frac13 \right)^4}{\Gamma\left(\frac23 \right)^2} 
\frac{-4\pi \left(51 i + 106 \sqrt{3}\right)  + 87 \left(5 + i \sqrt{3}\right) \log 27}{696(-5i+\sqrt{3})\pi^3}
 \\
\frac{\Gamma\left(\frac13 \right)^4}{\Gamma\left(\frac23 \right)^2} 
\frac{2\pi \left(-951 i + 137 \sqrt{3}\right)  + 609 \left(1 - i \sqrt{3}\right) \log 27}{696(-5i+\sqrt{3})\pi^3}
\end{array}\right) \;.
\ee
%\be
%\ba_{s,0}' = 
%\left( \begin{array}{c}  
%\frac{21\left(-3i+\sqrt{3}\right) \Gamma\left(\frac13 \right)^4}{4\left(-5i+\sqrt{3} \right)\pi^2\Gamma\left(\frac23 \right)^2} \\
%-\frac{3i\sqrt{3} \Gamma\left(\frac13 \right)^4}{4\pi^2\Gamma\left(\frac23 \right)^2}  \\
%-\frac{\Gamma\left(\frac13 \right)^4}{\Gamma\left(\frac23 \right)^2} 
%\frac{174(5+i\sqrt{3})\gamma +(465i+ 859\sqrt{3})\pi +174(5+i\sqrt{3})\left(2 \Gamma\left(0,\frac13 \right) - \Gamma\left(0,\frac23 \right) \right)}{696(-5i+\sqrt{3})\pi^3}
% \\
%\frac{\Gamma\left(\frac13 \right)^4}{\Gamma\left(\frac23 \right)^2} 
%\frac{1218i(i+\sqrt{3})\gamma -5(15i+ 67\sqrt{3})\pi +1218i(i+\sqrt{3})\left(2 \Gamma\left(0,\frac13 \right) - \Gamma\left(0,\frac23 \right) \right)}{696(-5i+\sqrt{3})\pi^3}
%\end{array}\right) \;.
%\ee

%%%%%%%%%%%%%%%%%%%%
\subsection{The BPS spectrum}
%%%%%%%%%%%%%%%%%%%%

To study the BPS states we can apply the general central charge expression (\ref{asympccgen}). However, in this case we should note that there is an additional factor of $z$ in the expression for the period vector, for example in (\ref{batyuz}), that can be removed by a Kahler transformation. More explicitly, we should utilise the expression 
\be
e^{\frac{K}{2}} = \frac{2 \pi^{\frac52}}{9\left|z\right|\sqrt{-\log \left|z \right|}} \left(\frac{\Gamma\left(\frac23 \right)}{\Gamma\left(\frac13 \right)^2}\right)^2 + ... \;.
\label{kahpota}
\ee
in the central charge formula. This yields the form (\ref{asympccgen}), up to a phase, after cancelling the factors of $|z|$ in (\ref{kahpota}) and in the period vector. The left over phase is universal to all the BPS states, so is independent of $\bq$. It therefore does not feature in any relative phases, which are the quantities of interest for us, and therefore we drop this phase from our expressions and work with (\ref{asympccgen}).

Since this is a $n=d=1$ locus, we can split the states into electric and magnetic as described in section \ref{sec:bpsmodst}.  

\subsubsection*{The electric states}

Electric states satisfy $\left<\bq_e,N \cdot \ba_0\right>=0$ and become massless in the degeneration limit. It is cleanest to give these after the symplectic rotation (\ref{Msym}), so we define
\be
\bq = M \cdot \bq' \;.
\ee
The electric states form a two-parameter family and are denoted as $\bq_{e}\left(r_1,r_2\right)$. They are given by \footnote{These states were first identified in \cite{Joshi:2019nzi} in a different symplectic basis.} 
\be
\mathrm{Electric\;} :\; \bq_{e}\left(r_1,r_2\right) = \frac{r_1}{3} \bx_1 + \frac{r_2}{3} \bx_2 \;, \;\; r_1,r_2 \in \mathbb{Z} \;,\; 4 r_1+ r_2  = 0 \; \mathrm{mod\;} 3\;,
\label{masslgen}
\ee
where
\be
\bx_1= \left( \begin{array}{c} 0 \\ 0 \\ 1 \\ 0 \end{array} \right) \;,\;\; \bx_2= \left( \begin{array}{c} 0 \\ 0 \\ 0 \\ 1 \end{array} \right) \;.
\ee
Note that the quantization is such that $M^{-1} \cdot \bq_{e}$ (and below $M^{-1} \cdot \bq_{m}$) is an integer vector.\footnote{It is not clear to the author that this is the correct quantization condition. For example, there are various strange looking factors of $\pi$ in some of the expressions. This may be related to the fact that the magnetic states are actually dyonic with respect to the gauge couplings. In any case, the exact quantization condition is not important for this work.}

%Importantly, the grade of massless states only has dependence on the monodromy angle through terms suppressed by $z$. Let us write 
%\be
%z = \left|z\right| e^{2 \pi i \theta} \;.
%\label{monthetdef}
%\ee
%Then we have
%\be
%Z\left(\bq'_e\right) = \frac{1}{\left| c \log z \right|^{\frac{1}{2}}} \left[ \left<\bq'_e, \ba'_0 \right> + \left|z\right| e^{2 \pi i \theta} \;\left(\bq'_0\right)^T \cdot \eta \cdot \ba'_1 + ... \right] \;.
%\label{asympcc0}
%\ee
%So the phase of the central charge, or the grade, for massless states is set by the constant $\left(\bq'_0\right)^T \cdot \eta \cdot \ba'_0$ up to corrections which behave as $|z|$ and so vanish near the degeneration locus
%\be
%\varphi\left(\bq'_0\right)  = \frac{1}{\pi} \mathrm{Im} \log \left[\left(\bq'_0\right)^T \cdot \eta \cdot \ba'_0\right]  + {\cal O}\left(|z|\right) \;.
%\ee

\subsubsection*{The Magnetic states}

The magnetic states form a two-parameter family and are denoted as $\bq_{m}\left(p_1,p_2\right)$. They are given by  
\be
\mathrm{Massive\;} :\; \bq_{m}\left(p_1,p_2\right) =  2 p_1 \bw_1 + 6 p_2 \bw_2 \;, \;\; p_1,p_2 \in \mathbb{Z} \;,
\ee
where
\be
\bw_1= \left( \begin{array}{c} 2 \\ 5 \\ 0 \\ 0 \end{array} \right) \;,\;\; \bw_2= \left( \begin{array}{c} 5 \\ -2 \\ 0 \\ 0 \end{array} \right) \;.
\ee

\subsubsection*{The central charge}

We can now evaluate explicitly the central charges as in (\ref{Ztd1n1}). This gives
\bea
Z\left(\bq,t\right) &=& \frac{1}{\left| c\; \mathrm{Im\;}t \right|^{\frac{1}{2}}} \left[ \left<\bq,\ba_0\right>  + \left<\bq,N\cdot\ba_0\right> t + ... \right] \;, \label{cexpr} \\
c &=& \frac{81}{8\pi^4}\left(\frac{\Gamma\left(\frac13 \right)^8}{\Gamma\left(\frac23 \right)^4}\right)\;, \\
 \left<\bq,\ba_0\right> &=& \frac{1}{4\left(-5i+\sqrt{3}\right)\pi^3}\left(\frac{\Gamma\left(\frac13 \right)^4}{\Gamma\left(\frac23 \right)^2} \right) \times  \\\nn
 & & \Bigg[
\pi \left( 6\left(-19i+\sqrt{3}\right)p_1 + \left(96i-92\sqrt{3} \right)p_2 - 7 \left(-3i+\sqrt{3}\right)  r_1 + \left(3i+5\sqrt{3}\right) r_2\right) \nn \\
& & + 9\log 3 \left( \left(15-11i\sqrt{3} \right)p_1 + \left(11+19i\sqrt{3} \right)p_2 \right) \Bigg]\;, \nn \\
%\left<\bq',\ba'_0\right> &=& -\frac{1}{4\left(-5i+\sqrt{3}\right)\pi^3}\left(\frac{\Gamma\left(\frac13 \right)^4}{\Gamma\left(\frac23 \right)^2} \right) \times  \\\nn
% & & \Bigg[
%6 \gamma \left( \left(15-11i\sqrt{3}\right)p_1 + \left(11+19i\sqrt{3} \right)p_2 \right) + 3\left(5i+13\sqrt{3}\right)\pi p_1 \nn \\
%& & + 25 \left(3i+5\sqrt{3}\right) \pi p_2 + 7 \left(-3i+\sqrt{3}\right) \pi r_1 - \left(3i+5\sqrt{3}\right) \pi r_2 \nn \\
%& & + 6 \left(\left(15-11i\sqrt{3} \right)p_1 + \left(11+19i\sqrt{3} \right)p_2\right) \left(2\Gamma\left(0,\frac13\right)- \Gamma\left(0,\frac23\right)\right) \Bigg]\;, \nn \\
 \left<\bq,N\cdot\ba_0\right> &=& 
 -\frac{9\Big[\left(15i+11\sqrt{3}\right)p_1+\left(11i-19\sqrt{3}\right)p_2  \Big]}{2\left(-5i+\sqrt{3}\right)\pi^2}\left(\frac{\Gamma\left(\frac13 \right)^4}{\Gamma\left(\frac23 \right)^2} \right) \;.
\label{Ztayex}
\eea
Here we wrote $\bq = \bq_e + \bq_m$, so that we have the integer magnetic charges $p_1, p_2$ and integer electric charges $r_1,r_2$.

%%%%%%%%%%%%%%%%%%%%
\subsection{Conifold and Large Complex Structure loci}
\label{sec:conlcslo}
%%%%%%%%%%%%%%%%%%%%

In this section we give the form of the period vector and BPS states for the conifold locus. We also give the monodromy matrix of the large complex structure.

%%%%%%%%%%%%%%%%%%%%
\subsubsection{The conifold locus}
\label{sec:conperiod}
%%%%%%%%%%%%%%%%%%%%

The period vector near the conifold locus was calculated in \cite{Joshi:2019nzi}. Expanding the Picard-Fuchs equations about the conifold locus gives the period vector
\be
\tilde{\bP}_C = \left( \begin{array}{c} 1+\frac{1}{243}u^3 + ... \\ u + \frac{13}{18} u^2+ ... \\ u^2 + \frac{67}{54} u^3 + .. \\ \left(u + \frac{13}{18} u^2 \right) \log u - \frac{323 u^3}{5832} + ... \end{array} \right) \;.
\label{conbasisper}
\ee
We need to transform this to a symplectic basis through
\be
\bP_C= U_{MC} \cdot \tilde{\bP}_C \;.
\ee
The explicit form of the transformation matrix is \cite{Joshi:2019nzi}
\be
U_{MC} = \left( \begin{array}{cccc} 
0 & -0.47746 i & 0 & 0 \\
7.2268 + 0.53267 i & 1.4974 + 0.074299 i & -0.22102 - 0.015039 i & 0 \\
1.0922 & -0.029110 & 0.006133 & 0.075991 \\
1.0653 i & 0.14859 i & -0.030072 i & 0
 \end{array}\right) \;.
\ee
This yields 
\be
\bP_C =  \left( \begin{array}{c} -2 \pi i \nu \; u + 0.344832 i u^2 + ...  \\ 
(7.2268 + 0.53267 i) + (1.4974 + 0.074299 i) u \\
1.0922 + \nu \;u \log u - 0.02911 u \\
1.0653 i + 0.14859 i u \\
\end{array} \right) \;,
\label{consymper}
\ee
with $\nu \simeq 0.07599$. From this we can determine the cycle that vanishes in the conifold limit $u \rightarrow 0$ as the top entry in the symplectic basis. 

By sending $u \rightarrow u e^{- 2 \pi i}$ in (\ref{conbasisper}) we can obtain the monodromy matrix about the conifold as 
\be
\tilde{N}_C = -2 \pi i\left( \begin{array}{cccc} 
0 & 0 & 0 & 0 \\
0 & 0 & 0 & 0 \\
0  & 0 & 0 & 0 \\
0 & 1  & 0 & 0 
\end{array}\right) \;.
\ee
We can translate this to the monodromy matrix in the symplectic basis by using (\ref{Tmat}) through
\be
N_C = U_{MC} \cdot \tilde{N}_C \cdot U_{MC}^{-1} \;,
\ee
which gives
\be
N_C = \left( \begin{array}{cccc} 
0 & 0 & 0 &0 \\
0 & 0 & 0 &0 \\
1 & 0 & 0 &0 \\
0 & 0  & 0 &0 
\end{array} \right)\;.
\ee

We can write the period vector in the canonical Nilpotent orbit form as
\be
\tilde{\bP}_C\left( u \right) = e^{-\frac{\tilde{N}_C \log u}{2\pi i}} \tilde{\ba}_C\left( u \right) \;.
\ee
The holomorphic vector takes the form
\be
\tilde{\ba}_C\left(u \right) =  \left( \tilde{\ba}_{C,0} + u \;\tilde{\ba}_{C,1} \right) + {\cal O}\left( u^{2}\right) \;,
\ee
with the expressions for the vectors
\be
\tilde{\ba}_{C,0} = \left( \begin{array}{c}  1 \\ 0 \\ 0 \\ 0 \end{array}\right) \;,\;\; 
\tilde{\ba}_{C,1} = \left( \begin{array}{c} 0 \\  1 \\ 0 \\ 0 \end{array}\right) 
\ee
We can also define the symplectic limiting vector
\be
\ba_C\left( u \right) \equiv U_{MC} \cdot \tilde{\ba}_C\left( u \right) \;,
\ee
and write
\be
\bP_C = e^{\frac{N_C \log u}{2\pi i}} \ba \left( u \right) \;.
\ee

%%%%%%%%%%%%%%%%%%%%
\subsubsection{The large complex structure locus}
%%%%%%%%%%%%%%%%%%%%

We will only be interested in the monodromy matrix about large complex structure, which in an appropriate symplectic basis takes the form \cite{Joshi:2019nzi}\footnote{Note that this is the inverse of the matrix in \cite{Joshi:2019nzi} which is given for a counter clockwise monodromy.}
\be
T_M = \left( \begin{array}{cccc} 
1 & 1 & -6 & 4 \\
0 & 1 & -5 & 9 \\
0 & 0 & 1 & 0 \\
0 & 0 & -1 & 1 
\end{array}\right) \;.
\label{tmlcs}
\ee

\bibliographystyle{jhep}
\bibliography{susyswamp.bib}  
\end{document}